\PassOptionsToPackage{unicode}{hyperref}
\PassOptionsToPackage{hyphens}{url}
\PassOptionsToPackage{dvipsnames,svgnames,x11names}{xcolor}

\documentclass[
  12pt]{article}

\usepackage{amsmath,amssymb}
\usepackage{iftex}
\ifPDFTeX
  \usepackage[T1]{fontenc}
  \usepackage[utf8]{inputenc}
  \usepackage{textcomp} 
\else 
  \usepackage{unicode-math}
  \defaultfontfeatures{Scale=MatchLowercase}
  \defaultfontfeatures[\rmfamily]{Ligatures=TeX,Scale=1}
\fi
\usepackage{lmodern}
\ifPDFTeX\else  
\fi
\IfFileExists{upquote.sty}{\usepackage{upquote}}{}
\IfFileExists{microtype.sty}{
  \usepackage[]{microtype}
  \UseMicrotypeSet[protrusion]{basicmath} 
}{}
\makeatletter
\@ifundefined{KOMAClassName}{
  \IfFileExists{parskip.sty}{%
    \usepackage{parskip}
  }{
    \setlength{\parindent}{0pt}
    \setlength{\parskip}{6pt plus 2pt minus 1pt}}
}{
  \KOMAoptions{parskip=half}}
\makeatother
\usepackage{xcolor}
\makeatletter
\ifx\paragraph\undefined\else
  \let\oldparagraph\paragraph
  \renewcommand{\paragraph}{
    \@ifstar
      \xxxParagraphStar
      \xxxParagraphNoStar
  }
  \newcommand{\xxxParagraphStar}[1]{\oldparagraph*{#1}\mbox{}}
  \newcommand{\xxxParagraphNoStar}[1]{\oldparagraph{#1}\mbox{}}
\fi
\ifx\subparagraph\undefined\else
  \let\oldsubparagraph\subparagraph
  \renewcommand{\subparagraph}{
    \@ifstar
      \xxxSubParagraphStar
      \xxxSubParagraphNoStar
  }
  \newcommand{\xxxSubParagraphStar}[1]{\oldsubparagraph*{#1}\mbox{}}
  \newcommand{\xxxSubParagraphNoStar}[1]{\oldsubparagraph{#1}\mbox{}}
\fi
\makeatother

\usepackage{longtable,booktabs,array}
\usepackage{calc} 
\usepackage{etoolbox}
\makeatletter
\patchcmd\longtable{\par}{\if@noskipsec\mbox{}\fi\par}{}{}
\makeatother
\IfFileExists{footnotehyper.sty}{\usepackage{footnotehyper}}{\usepackage{footnote}}
\makesavenoteenv{longtable}
\usepackage{graphicx}
\makeatletter
\def\maxwidth{\ifdim\Gin@nat@width>\linewidth\linewidth\else\Gin@nat@width\fi}
\def\maxheight{\ifdim\Gin@nat@height>\textheight\textheight\else\Gin@nat@height\fi}
\makeatother
\setkeys{Gin}{width=\maxwidth,height=\maxheight,keepaspectratio}
\makeatletter
\def\fps@figure{htbp}
\makeatother

\makeatletter
\@ifpackageloaded{caption}{}{\usepackage{caption}}
\AtBeginDocument{%
\ifdefined\contentsname
  \renewcommand*\contentsname{Table of contents}
\else
  \newcommand\contentsname{Table of contents}
\fi
\ifdefined\listfigurename
  \renewcommand*\listfigurename{List of Figures}
\else
  \newcommand\listfigurename{List of Figures}
\fi
\ifdefined\listtablename
  \renewcommand*\listtablename{List of Tables}
\else
  \newcommand\listtablename{List of Tables}
\fi
\ifdefined\figurename
  \renewcommand*\figurename{Figure}
\else
  \newcommand\figurename{Figure}
\fi
\ifdefined\tablename
  \renewcommand*\tablename{Table}
\else
  \newcommand\tablename{Table}
\fi
}
\@ifpackageloaded{float}{}{\usepackage{float}}
\floatstyle{ruled}
\@ifundefined{c@chapter}{\newfloat{codelisting}{h}{lop}}{\newfloat{codelisting}{h}{lop}[chapter]}
\floatname{codelisting}{Listing}

\makeatother
\makeatletter
\@ifpackageloaded{caption}{}{\usepackage{caption}}
\@ifpackageloaded{subcaption}{}{\usepackage{subcaption}}
\makeatother

\ifLuaTeX
  \usepackage{selnolig}  
\fi
\usepackage[]{natbib}
\usepackage{bookmark}

\IfFileExists{xurl.sty}{\usepackage{xurl}}{} 
\hypersetup{
  pdftitle={Title},
  pdfauthor={Author 1; Author 2},
  pdfkeywords={3 to 6 keywords, that do not appear in the title},
  colorlinks=true,
  linkcolor={blue},
  filecolor={Maroon},
  citecolor={Blue},
  urlcolor={Blue},
  pdfcreator={LaTeX via pandoc}}

\newcommand{\anon}{1}

\newcommand{\E}{\mbox{$\mathbb{E}$}}
\newcommand{\R}{\mbox{$\mathbb{R}$}}
\newcommand{\bR}{\mbox{$\mathbf{R}$}}
\newcommand{\bX}{\mbox{$\mathbf{X}$}}
\newcommand{\bT}{\mbox{$\mathbf{T}$}}
\newcommand{\bY}{\mbox{$\mathbf{Y}$}}
\DeclareMathOperator*{\argmin}{arg\,min}
\newcommand{\norm}[1]{\left \lVert #1 \right \rVert}
\usepackage{threeparttable}
\usepackage{booktabs} 
\usepackage{graphicx}
\usepackage{mathtools}  
\usepackage{multirow}
\usepackage{amsthm}
\usepackage{tikz}
\usetikzlibrary{
  arrows.meta, positioning, calc, fit,
  backgrounds, decorations.pathreplacing
}

\newtheorem{theorem}{Theorem}[section]
\newtheorem{corollary}[theorem]{Corollary}

\newtheorem{remark}[theorem]{Remark}

\usepackage{algorithm}
\usepackage{algorithmic}
\usepackage{float}

\let\hat\widehat
\let\tilde\widetilde

\begin{document}

\def\spacingset#1{\renewcommand{\baselinestretch}%
{#1}\small\normalsize} \spacingset{1}


\if1\anon
{
  \title{\bf Emputation: Identification-Guided Neural Imputation Framework}
  \author{Yanjiao Yang, Yikun Zhang, Xinwei Shen, Yen-Chi Chen\thanks{
    The authors gratefully acknowledge \textit{NSF Grant DMS-2141808 (CAREER), 2310578 and NIH Grant U24 AG072122}}\hspace{.2cm}\\
    Department of Statistics, University of Washington\\
    }
  \maketitle
} \fi

\if0\anon
{
  \bigskip
  \bigskip
  \bigskip
  \begin{center}
    {\LARGE\bf Title}
\end{center}
  \medskip
} \fi

\bigskip
\begin{abstract}
We propose Emputation, a deep generative framework for learning imputation models. Emputation targets the extrapolation distribution of missing variables given observed variables, and training is guided by specific missingness assumptions that guarantee identification of the target distribution. The training objective, called the emputation risk, is an energy-score-based risk in which the identification assumption determines how observed entries are masked and which observations contribute to training. The resulting framework enables direct conditional sampling for multiple imputation. 
We show that the population minimizer of the emputation risk recovers the target extrapolation distribution under a broad class of identification assumptions, including several missing-not-at-random assumptions. Simulations show strong performance under both pointwise and distributional evaluation metrics, and an application to an Alzheimer's disease dataset demonstrates its practical value.
\end{abstract}

\noindent%
{\it Keywords:} generative models, pattern-mixture models, energy score, multiple imputation, missing-not-at-random
\vfill

\newpage
\spacingset{1.8} 

\section{Introduction}\label{sec-intro}

Missing data are prevalent in real-world datasets and arise in a wide range of scientific domains \citep{Hirano2001, vansteelandt2006ignorance, sterne2009multiple}. If handled improperly, missingness can lead to biased estimation, loss of efficiency, and unreliable downstream analyses \citep{little2019statistical}. 
Imputation is a popular strategy for handling missing data \citep{vanbuuren, rubin2018multiple,carpenter2023multiple} because once a complete dataset is obtained, downstream analyses can proceed using conventional methods.
The idea of imputation is simple: one trains a model that (stochastically) fills in the missing entries according to both a missing data assumption, which identifies the distribution of the missing values from the observed data, and a modeling assumption, which specifies how this distribution is estimated. In the statistics community, an imputation model is typically constructed by choosing a reasonable missing data assumption and a relatively simple model. This approach often leads to an interpretable imputation model \citep{van1999flexible,molenberghs1998missing,daniels2023bayesian}, but it may not capture the complexity of the underlying distribution. In contrast, the modern machine learning community focuses on training deep generative models that can approximate the underlying distribution accurately. However, the training objectives in these models often do not explicitly encode the identifying missing data assumption, and many empirical evaluations are conducted under missing completely at random or related masking schemes \citep{yoon2018gain,misGAN}. 
In this paper, we aim to bridge the gap between these two communities. We propose \emph{emputation}, a deep generative modeling framework where the training objective is guided by a missing data assumption in the form of a pattern mixture model \citep{little1993pattern, daniels2008missing}. 

\subsection{Overview of Emputation}
We now provide an overview of our proposed framework. The name \emph{emputation} combines \emph{engression} \citep{shen2025engression} and \emph{imputation}. In imputation, missing values are replaced by draws from a statistical model, enabling standard complete-data workflows to be applied. Engression, which we review in Section~\ref{sec::emputation}, is a neural network-based generative modeling framework for distributional regression based on the energy score. 

Emputation is formulated as an empirical risk minimization problem, where the training objective (which we call the \emph{emputation risk}) is built on three ingredients: (1) the energy score, which targets distributional learning; (2) a masking strategy, where observed entries are strategically masked and reconstructed during training; and (3) a selection criterion, which selects a subset of observations for masking according to their response patterns. Once the neural network is trained by minimizing the emputation risk, missing entries can be imputed directly by sampling from the learned model. 
What distinguishes emputation from many existing deep generative imputation methods is that the emputation risk is not merely a heuristic training objective, but is guided by a chosen missing data assumption. This is achieved by masking and selecting the right target variables and individuals.

\subsection{Related Work}
Since the literature on missing data is vast, we focus our review on two lines of work most relevant to our emputation framework: pattern mixture models and deep generative imputation.

\paragraph*{Pattern mixture models.}
Pattern mixture (PM) models factorize the joint distribution of the data and response pattern into the marginal distribution of the missingness indicators and the pattern-specific data distribution \citep{little1993pattern}. Because the extrapolation density (\emph{i.e.}, the conditional density of missing variables given the observed ones and response pattern) is unidentifiable, identification requires restrictions that relate missing components to observed variables. Notable examples include complete-case missing value (CCMV; \citealt{little1993pattern, discreteEric2018}), available-case missing value (ACMV; \citealt{molenberghs1998missing}), nearest-case missing value (NCMV; \citealt{daniels2008missing}) assumptions, or more general pattern-borrowing restrictions \citep{daniels2023bayesian} and pattern graphs \citep{chen2022pattern}. Existing work has developed identification theory and Bayesian inference under these restrictions \citep{thijs2002strategies, linero2018bayesian}. Nevertheless, less attention has been given to how these identifying restrictions within PM models can be translated into scalable training objectives of deep generative models for imputation.

\paragraph*{Deep generative models for imputation.} 
Recent imputation methods have increasingly adopted deep generative models to handle complex dependence structures in large data, including variational autoencoder (VAE)-based methods \citep{miwae, nazabal2020handling}, generative adversarial network (GAN)-based methods \citep{yoon2018gain, misGAN}, and diffusion-based methods \citep{CSDI, ouyang2023missdiff, zhang2025diffputer}. Much of this literature is developed under missing completely at random (MCAR) or missing at random (MAR) assumptions, although recent work has extended deep generative imputation to missing not at random (MNAR) settings through latent variable models with variational inference \citep{ipsen2021notmiwae, ma2021identifiable, Ghalebikesabi2021}. Despite this progress, direct conditional sampling of missing values given observed entries is often nontrivial \citep{miwae, richardson2020mcflow}, and imputation performance is most commonly evaluated through pointwise accuracy metrics such as RMSE, with limited attention to distributional alignment.

\section{Emputation}
\label{sec::emputation}

\subsection{Background and Notation}
Let $X \in \mathbb{R}^d$ denote the study variables subject to missingness and $R \in \{0,1\}^d$ denote the response vector of $X$, where $R_j = 1$ if $X_j$ is observed and $R_j = 0$ otherwise. Let $\bar{R} = 1_d - R$ be the complement (flipping 0 and 1) of pattern $R$, where $1_d$ is the complete-case response pattern. For notational convenience, we write $R=r$ as shorthand for $(R_1,\dots,R_d)=(r_1,\dots,r_d)$.
We use $X_r = (X_j : r_j = 1)$ to denote the observed entries under response pattern $R=r$, and $X_{\bar{r}}$ to denote the missing variables under that pattern. With this notation, an observation can be expressed as $(X_R,R)$. We assume the observed data consist of $n$ independent and identically distributed (i.i.d.) observations
$\{(\bX_{i, \bR_i}, \bR_i)\}_{i=1}^n$. 
The boldface notation, such as $\bX_i \in \R^d$ and $\bR_i\in\{0,1\}^d$, is used for highlighting the $i$-th observation in the observed data. We use $I(\cdot)$ to denote the indicator function.

We further define the size of pattern $r$ as $|r|=\sum_{j=1}^dr_j$. In particular, the size of the complete-case pattern is $d$, i.e., $|1_d|=d$.
For two response patterns $r_1, r_2$, we define $r_1>r_2$ if $r_{1,j}\geq r_{2,j}$ for all $j\in \{1,\dots,d\}$, with strict inequality for at least one coordinate. Thus, when $r_1>r_2$, the variables observed under pattern $r_2$ form a strict subset of those observed under pattern $r_1$.
Let $\mathcal{R}=\{r\neq 1_d: p(R=r)>0\}$ denote the collection of incomplete response patterns with positive probability, and assume $p(R=1_d|X)>0$ almost surely.

The probability model for missing data can be characterized through the joint distribution of $(X,R)$.  
Under the pattern mixture model factorization \citep{little1993pattern}, the joint distribution $p(x,r)$ decomposes into $p(x,r) = p(x_{\bar r}|x_r,r) p(x_r,r)$. 
Using the above notation, $x_{\bar r}$ refers to missing variables and $x_r$ refers to observed variables under response pattern $R=r$.
Here, $p(x_r,r)$ is the observed-data distribution and is identifiable from the data, whereas $p(x_{\bar r}|x_r,r)$ is known as \emph{the extrapolation density/distribution} \citep{little1993pattern, daniels2008missing} characterizing 
the conditional distribution of the missing components given the observed components under the response pattern $R=r$. 
This extrapolation density is not identifiable from the observed data without additional assumptions. 

An imputation model can be viewed as a model for the extrapolation density $p(x_{\bar r}|x_r,r)$. From the distributional perspective, an imputation model is inherently generative, because once the conditional distribution is learned, missing values can be generated by sampling from the fitted model.

\subsection{Engression}
Engression \citep{shen2025engression} is a generative modeling framework for distributional regression that extends classical regression from estimating the conditional mean to learning the conditional distribution of response $Y\in\mathbb{R}^q$ given covariates $X\in\mathbb{R}^d$. The learning procedure starts with modeling $Y=f(X,\epsilon)$, where $\epsilon\in\mathbb{R}^q$ is a random noise vector from a fixed distribution such as $N(0,\mathbf{I}_q)$ with identity matrix $\mathbf{I}_q\in\mathbb{R}^{q\times q}$ and $f$ is parameterized by a neural network. 
In particular, $f$ is trained using the energy score, a strictly proper scoring rule \citep{gneiting2007strictly}. For a distribution $P$ and observation $y$, the energy score is
\begin{equation}
\text{ES}(P,y) = \frac{1}{2}\E\norm{Y-Y'} - \E\norm{Y-y},
\label{eq::energyscore}
\end{equation}
where $Y$ and $Y'$ are independent draws from $P$. The second term encourages the generated samples to stay close to the observed responses, while the first term prevents distributional collapse. At the population level, engression minimizes the expected negative energy score:
\begin{equation}
\begin{aligned}
L(f) 
=\E\left[ \norm{f(X,\epsilon)-Y} - \frac{1}{2}\norm{f(X,\epsilon)-f(X,\epsilon')} \right],
\label{eq::engression_ppl}
\end{aligned}
\end{equation}
where $\epsilon,\epsilon'\sim N(0,\mathbf{I}_q)$.
Given an i.i.d.\ sample $\{(\bX_i,\bY_i)\}_{i=1}^n$ and independent noise draws $\{\epsilon_{i,b}\sim N(0,\mathbf{I}_q)\}_{b=1}^B$, the empirical risk of engression is 
\begin{equation}
\begin{aligned}
\hat{L}(f) &=\frac{1}{n}\sum_{i=1}^n \left[\frac{1}{B}\sum_{b=1}^B\norm{\bY_i-f(\bX_i,\epsilon_{i,b})} - \frac{1}{2B(B-1)}\sum_{b, b'=1}^B\norm{f(\bX_i,\epsilon_{i,b}) - f(\bX_i,\epsilon_{i,b'})}
\right],
\label{eq::engression_sample}
\end{aligned}
\end{equation}
which can be optimized using standard optimization methods.
Here, $B$ is the number of Monte Carlo draws used to approximate the population risk.
Once a model $\hat f$ has been trained, generating $Y$ from any given covariate $x$ can be done by $\hat Y = \hat f(x, \epsilon)$, where $\epsilon \sim N(0,\mathbf{I}_q)$.

The high-level idea of how engression works is apparent in the one-dimensional case. When $Y\in\mathbb{R}$, 
the minimizer of the population $L(f)$ in \eqref{eq::engression_ppl} is the inverse conditional cumulative distribution function (CDF) of $Y$ given $X=x$:
$$
f^*(x,\epsilon) = F^{-1}_{Y|X}(\Phi(\epsilon)\mid x),
$$
where $\Phi$ is the CDF of $N(0,1)$ so that $\Phi(\epsilon)$ is a uniform random variable over $[0,1]$. 
Thus, the estimator $\hat f$ can be viewed as an approximation of the conditional inverse CDF with a neural network. 

\subsection{Emputation Methodology} \label{sec::emp::method}
Emputation is an imputation framework that learns the extrapolation density by combining engression 
and missing data assumptions in the form of a pattern mixture model.  

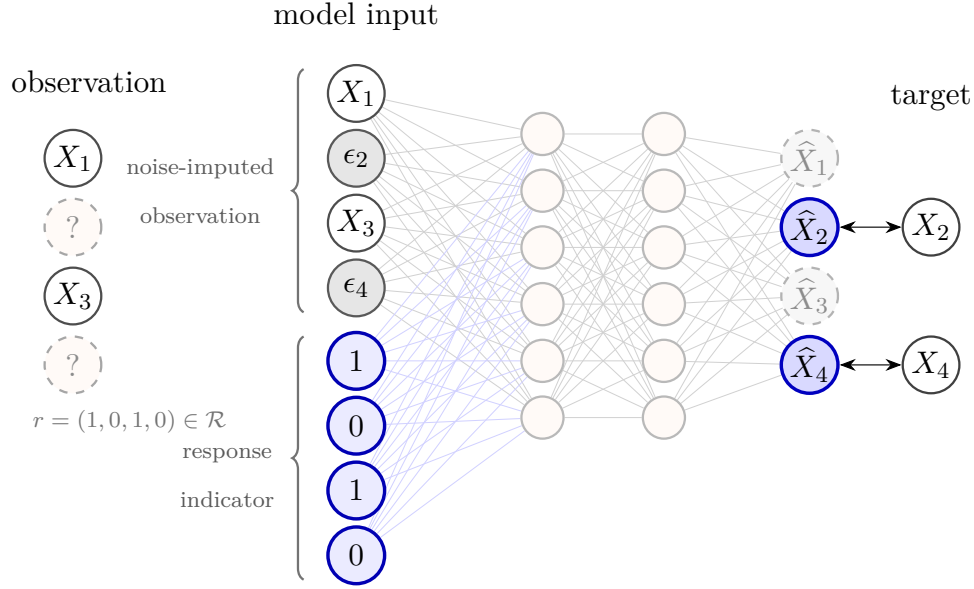
\begin{figure}[t]
\centering

\definecolor{softpeach}{RGB}{255,250,247}

\begin{subfigure}{\textwidth}
\centering
\vspace{6pt}
\resizebox{0.8\textwidth}{!}{%
\begin{tikzpicture}[
  font=\small,
  >={Stealth[length=2mm]},
  obs/.style    ={circle,draw=black!70,thick,minimum size=7mm,inner sep=0pt,fill=white},
  qmark/.style  ={circle,draw=black!35,thick,dashed,minimum size=7mm,inner sep=0pt,
                  fill=softpeach,text=black!45},
  eps/.style    ={circle,draw=black!65,thick,minimum size=7mm,inner sep=0pt,
                  fill=black!10},
  pat/.style    ={circle,draw=blue!70!black,very thick,minimum size=7mm,inner sep=0pt,
                  fill=blue!8},
  hid/.style    ={circle,draw=gray!55,thick,minimum size=5.2mm,inner sep=0pt,
                  fill=softpeach},
  outign/.style ={circle,draw=black!30,thick,dashed,minimum size=7mm,inner sep=0pt,
                  fill=black!3,text=black!45},
  targ/.style   ={circle,draw=blue!75!black,very thick,minimum size=7mm,inner sep=0pt,
                  fill=blue!15},
  truth/.style  ={circle,draw=black!75,thick,minimum size=7mm,inner sep=0pt,
                  fill=white},
  conn/.style   ={draw=black!18,line width=0.25pt},
  pconn/.style  ={draw=blue!18,line width=0.25pt}
]

\node[font=\small] at (-5.55, 2.2) {observation};
\node[font=\small] at (-2.25, 3) {model input};
\node[font=\small] at (4.85, 2) {target};

\node[obs]   (R1) at (-5.75,  1.25) {$X_1$};
\node[qmark] (R2) at (-5.75,  0.40) {$?$};
\node[obs]   (R3) at (-5.75, -0.45) {$X_3$};
\node[qmark] (R4) at (-5.75, -1.30) {$?$};

\node[align=center,font=\scriptsize,text=black!55] at (-5.05,-2.00)
{$r=(1,0,1,0)\in\mathcal{R}$};

\node[obs] (I1) at (-2.25,  2.05) {$X_1$};
\node[eps] (I2) at (-2.25,  1.25) {$\epsilon_2$};
\node[obs] (I3) at (-2.25,  0.45) {$X_3$};
\node[eps] (I4) at (-2.25, -0.35) {$\epsilon_4$};

\node[pat] (P1) at (-2.25, -1.25) {$1$};
\node[pat] (P2) at (-2.25, -2.05) {$0$};
\node[pat] (P3) at (-2.25, -2.85) {$1$};
\node[pat] (P4) at (-2.25, -3.65) {$0$};

\draw[decorate,decoration={brace,mirror,amplitude=4pt},draw=black!55,thick]
  (-2.90,2.35) -- (-2.90,-0.65)
  node[midway,left=6pt,align=center,font=\scriptsize,text=black!65]
  {noise-imputed\\observation};

\draw[decorate,decoration={brace,mirror,amplitude=4pt},draw=black!55,thick]
  (-2.90,-0.95) -- (-2.90,-3.95)
  node[midway,left=6pt,align=center,font=\scriptsize,text=black!65]
  {\\response\\indicator};

\foreach \k/\y in {1/1.55, 2/0.85, 3/0.15, 4/-0.55, 5/-1.25, 6/-1.95}
  \node[hid] (H1\k) at (0.05,\y) {};

\foreach \k/\y in {1/1.55, 2/0.85, 3/0.15, 4/-0.55, 5/-1.25, 6/-1.95}
  \node[hid] (H2\k) at (1.55,\y) {};

\node[outign] (O1) at (3.35,  1.25) {$\hat X_1$};
\node[targ]   (O2) at (3.35,  0.40) {$\hat X_2$};
\node[outign] (O3) at (3.35, -0.45) {$\hat X_3$};
\node[targ]   (O4) at (3.35, -1.30) {$\hat X_4$};

\node[truth] (T2) at (4.85,  0.40) {$X_2$};
\node[truth] (T4) at (4.85, -1.30) {$X_4$};

\draw[<->] (O2) -- (T2);
\draw[<->] (O4) -- (T4);

\begin{scope}[on background layer]
  \foreach \s in {I1,I2,I3,I4}
    \foreach \t in {1,...,6}
      \draw[conn] (\s) -- (H1\t);

  \foreach \s in {P1,P2,P3,P4}
    \foreach \t in {1,...,6}
      \draw[pconn] (\s) -- (H1\t);

  \foreach \s in {1,...,6}
    \foreach \t in {1,...,6}
      \draw[conn] (H1\s) -- (H2\t);

  \foreach \s in {1,...,6}
    \foreach \t in {1,...,4}
      \draw[conn] (H2\s) -- (O\t);
\end{scope}

\end{tikzpicture}%
}
\end{subfigure}

\caption{Emputation model architecture. For $r=(1,0,1,0)\in\mathcal{R}$, observed coordinates $X_1$ and $X_3$ are kept in their original positions, while the missing coordinates are replaced by independent noise variables $\epsilon_2$ and $\epsilon_4$. The resulting noise-imputed observation and the response indicator $r=(1,0,1,0)$ are concatenated and then passed through the neural network. Only the imputed outputs corresponding to the missing entries under $r$, namely $\hat X_2$ and $\hat X_4$, are kept.}
\label{fig:emputation-framework}
\end{figure}

\paragraph*{Emputation model.} Consider a model class $\mathcal{F}$ consisting of vector-valued functions $f:\mathbb{R}^d \times \{0,1\}^d \to \mathbb{R}^d$. Each function $(x,r)\mapsto f(x,r)$ takes as input a completed $d$-dimensional vector $x$ and a response pattern $r$. 
For an incomplete observation $(X_R, R) = (x_r,r)$ with $r\neq 1_d$, we construct an input vector for $f$ by keeping the observed components $x_r$ in their original coordinates and replacing the missing components by random noise $\epsilon_{\bar r} \sim N(0, \mathbf{I}_{|\bar r|})$, and then appending the pattern $r$. For example, if $d=4$ and $r=(1,0,1,0)$, the input vector is
$(x_r,\epsilon_{\bar r},r)=(x_1,\epsilon_2,x_3,\epsilon_4, 1,0,1,0)$, where $\epsilon_2$ and $\epsilon_4$ are independent draws from the standard Gaussian distribution. \autoref{fig:emputation-framework} illustrates this construction.
Throughout the paper, we use Gaussian noise as the default reference distribution, although other fixed noise distributions, such as the uniform distribution, are applicable as well. 
The missing components in the incomplete observation are then imputed by 
$
\hat X_{\bar r} = [f(x_r, \epsilon_{\bar r}, r)]_{\bar r},
$ 
where $[f(z)]_{\bar r} = (f_j(z): r_j = 0)$ denotes the coordinates corresponding to the missing entries under pattern $r$. In other words, we impute the missing variables $x_{\bar r}$ by the random vector $\hat X_{\bar r}$ similar to how engression generates a new response variable.
Thus, for each $f$, the random output $\hat X_{\bar r}$ induces a conditional distribution (or density), denoted by $p_f(\cdot|x_r,r)$. 

In practice, to make the model class $\mathcal{F}$ operational, 
we constrain $\mathcal{F}$ to a class of functions
parameterized by a neural network. 
Specifically, each $f\in\mathcal F$ is written as $f_\theta(z)$ with
\begin{equation}
f_\theta(z)
=
h_L \circ \sigma \circ h_{L-1} \circ \cdots \circ \sigma \circ h_1(z),
\label{eq::NN}
\end{equation}
where $z=(x_r,\epsilon_{\bar r},r)$, $h_\ell(u)=W_\ell u+\eta_\ell$ is an affine map, with
$h_1:\mathbb R^{d}\times\{0,1\}^d\to\mathbb R^{q_1}$,
$h_\ell:\mathbb R^{q_{\ell-1}}\to\mathbb R^{q_\ell}$ for $\ell=2,\ldots,L-1$,
and $h_L:\mathbb R^{q_{L-1}}\to\mathbb R^d$.
The parameter $\theta=(W_1,\ldots,W_L,\eta_1,\ldots,\eta_L)$
includes all weight matrices and bias vectors.
The activation function $\sigma(\cdot)$ is applied componentwise and a typical choice is the ReLU activation function
$\sigma(t)=\max\{0,t\}$.

Once $f$ is parameterized as $f_\theta$, learning $f$ amounts to estimating the underlying parameter $\theta$, and the fitted imputation model is the neural network model $\hat f = f_{\hat \theta}$. 
Note that $\hat f$ is a single neural network model that works for the extrapolation density $p(x_{\bar r}|x_r,r)$ across every response pattern $r$. The key is to include the pattern vector $r$ as part of the input to the neural network and properly construct the masked pattern and selection function  according to the target missing data assumption, as described below.

\paragraph*{Emputation risk.} 
The key of emputation is to construct a risk function whose population minimizer $f^*$ satisfies $p_{f^*}(x_{\bar r}| x_r, r) = p(x_{\bar r}|x_r,r)$
under the chosen missing data assumption. Given i.i.d.\ observations $\{\bX_i,\bR_i\}_{i=1}^n$, we train $f$ by the following \emph{empirical emputation risk minimization}
\begin{equation}
\label{eq::emploss_pop}
\hat{f} = \argmin_{f\in\mathcal{F}}\hat{\mathcal{L}}(f) \quad \text{ with } \quad \hat{\mathcal{L}}(f)= \frac{1}{n}\sum_{i=1}^n\ell(f|\bX_{i,\bR_i},\bR_i)
\end{equation}
where the loss for observation $i$ is defined by summing the approximated (negative) energy score losses over incomplete response patterns $r\in \mathcal{R}$:
\begin{equation}
\begin{aligned}
\ell(f|\bX_{i,\bR_i},\bR_i)
&= \sum_{r\in\mathcal{R}}S_r(\bR_i)
\Bigg\{
\frac{1}{B}\sum_{b=1}^B \left\|
\bX_{i,M_r(\bR_i)} - [f(\bX_{i,r}, \epsilon_{i,\bar{r}, b}, r)]_{M_r(\bR_i)}\right\| \\
-\frac{1}{2B(B-1)}&\sum_{b, b'=1}^B \left\|[f(\bX_{i,r}, \epsilon_{i,\bar{r}, b}, r)]_{M_r(\bR_i)}  - [f(\bX_{i,r}, \epsilon_{i,\bar{r}, b'},r)]_{M_r(\bR_i)}\right\|  
\Bigg\}.
\label{eq::emploss}
\end{aligned}
\end{equation}
Here, for each target pattern $r$, the loss uses two key ingredients that incorporate the knowledge of a missing data assumption: 
the \emph{masked pattern} $M_r(\bR_i)\in \{0,1\}^d$ specifies which observed coordinates of observation $i$ are temporarily treated as targets to be reconstructed;
the \emph{selection function} $S_r(\bR_i) \in \mathbb{R}_{\geq 0}$ determines whether, and with what weight, observation $i$ contributes to the loss for pattern $r$. 
The noise vectors $\epsilon_{i,\bar{r}, b} \in\R^{|\bar r|}$ are independent draws from $N(0, \mathbf{I}_{|\bar r|})$ for each $i$ and $b$. 

The key distinction between emputation and a generic imputation method is that the masked pattern $M_r$ and the selection function $S_r$ are not arbitrary design choices. Instead, they encode the identifying restriction imposed by the chosen missing data assumption. Different assumptions lead to different choices of $M_r$ and $S_r$. Moreover, they are dependent in the sense that whenever $S_r(\bR_i)>0$, the coordinates selected by $M_r(\bR_i)$ need to be observed for observation $i$.

\begin{table}[t]
\centering
\caption{Masked pattern $M_r(\bR_i)$ and selection function $S_r(\bR_i)$ under different missing data assumptions.}
\label{tab::emp_summary}
\resizebox{\linewidth}{!}{%
\begin{tabular}{lllcc}
\toprule
& Section & Assumption & Masked pattern $M_r(\bR_i)$ & Selection function $S_r(\bR_i)$ \\
\midrule
\multirow{3}{*}{Nonmonotone} 
& Section~\ref{sec::mcar} & MCAR 
& $\bR_i-r$ 
& $|\bR_i-r|^{-1}I(r<\bR_i)$ \\
& Section~\ref{sec::ccmv} & CCMV 
& $\bar r$ 
& $|\bar r|^{-1}I(\bR_i=1_d)$ \\
& Section~\ref{sec::tree} & Tree graph 
& ${\sf PA}(r)-r$ 
& $|{\sf PA}(r)-r|^{-1}I(\bR_i={\sf PA}(r))$ \\
\midrule
\multirow{3}{*}{Monotone} 
& Section~\ref{sec::monotone_acmv} & m-ACMV 
& $|r|+1$ 
& $I(|\bR_i|\geq |r|+1)$ \\
& Section~\ref{sec::monotone_ccmv_ncmv} & m-CCMV 
& $|r|+1$ 
& $I(|\bR_i|=d)$ \\
& Section~\ref{sec::monotone_ccmv_ncmv} & m-NCMV 
& $|r|+1$ 
& $I(|\bR_i|=|r|+1)$ \\
\bottomrule
\end{tabular}
}
\end{table}

\autoref{tab::emp_summary} summarizes the choices of $M_r$ and $S_r$ for the missing data assumptions considered in this paper, which are detailed in the following sections. 
In the pattern mixture model framework, the masked pattern and selection function are constructed by how the extrapolation density is identified. For instance, under the CCMV assumption, the extrapolation density is identified using complete cases ($R=1_d$). Thus, for a target pattern $r$, we mask the variables in complete cases that would be missing under pattern $r$, corresponding to $M_r(\bR_i) = 1_d - r = \bar{r}$. The selection function involves the complete-case indicator $I(\bR_i = 1_d)$, along with an inverse weight for the number of masked variables, leading to $S_r(\bR_i ) = \frac{1}{|\bar r|} I(\bR_i = 1_d)$. See Section~\ref{sec::ccmv} for more details.

\begin{remark}
In this paper, we use neural networks to parameterize the emputation model because of the flexibility and expressiveness. The proposed framework, however, is not restricted to neural networks. More generally, the emputation framework can be implemented with other flexible distributional learners that induce a conditional distribution for the missing variables and support sampling from that distribution. Nonparametric methods, for instance, kernel methods or tree-based methods may also be incorporated with suitable modifications. In such extensions, the masked pattern and selection function would still be determined by the identification assumption. Therefore, the framework would target the same extrapolation density, while the choice of learner would only affect how this density is estimated in practice.
\end{remark}

\begin{remark}
The emputation risk is built on the energy score, reflecting our view of imputation as a distributional learning problem rather than merely a prediction task, as also emphasized by \citet{vanbuuren, naf2024good}. In particular, minimizing a pointwise prediction loss, such as mean squared error, generally targets a conditional mean \citep{yang2025masking}, whereas a good imputation method should be ``distributional or stochastic'' \citep{naf2026practicalguidemodernimputation}. In this paper, we focus on the energy score and develop the emputation framework under this choice. Extensions to other scoring rules are left for future work, as discussed in Section~\ref{sec::conclusion}.
\end{remark}

\section{Emputation Under Nonmonotone Missingness}
\label{sec:nonmon_emputation}
Nonmonotone missingness is a common and general scenario in missing data problems, where variables may be missing in arbitrary patterns. In this section, we illustrate how emputation applies to several well-established assumptions for nonmonotone missingness, including the MCAR and CCMV assumptions and tree graph restrictions \citep{suen2026modelingmultivariatemissingnesstree}.

\subsection{Emputation under MCAR}
\label{sec::mcar}
MCAR assumes that the missingness $R$ is independent of the data $X$, namely, $R\perp X$.
Under MCAR, we write the extrapolation density as 
\begin{equation*}
p(x_{\bar r}|x_r, r) = p(x_{\bar r}|x_r, r') = p(x_{\bar r}|x_r)
\end{equation*}
for any response patterns $r$ and $r'$ with positive probability. 
In particular, for patterns with more observed entries $r'>r$, MCAR implies the identified (partial) extrapolation density:
\begin{equation}
p(x_{r'- r}|x_r, r) = p(x_{r'- r}|x_r, r').
\label{eq::mcar2}
\end{equation}
The above equation suggests that, to train the imputation model $p(x_{\bar r}|x_r,r)$, we can use any observation with response pattern $\bR_i>r$ and mask the additionally observed variables $x_{\bR_i-r}$ when constructing the loss.

As a result, under MCAR, our masked pattern is $M_r(\bR_i)=\bR_i-r$
and the selection function $S_r(\bR_i)=I(r<\bR_i)/|\bR_i-r|$. 
The empirical objective function is
\begin{equation}
\begin{aligned}
\hat{\mathcal{L}}(f)
&= \frac{1}{n}\sum_{i=1}^n\Bigg[
\sum_{r<\bR_i} \frac{1}{|\bR_i-r|}
\Bigg\{
\frac{1}{B}\sum_{b=1}^B \left\|
\bX_{i,\bR_i-r} - [f(\bX_{i,r}, \epsilon_{i,\bar{r}, b}, r)]_{\bR_i-r}\right\| \\
&-\frac{1}{2B(B-1)}\sum_{b, b'=1}^B \left\|[f(\bX_{i,r}, \epsilon_{i,\bar{r}, b}, r)]_{\bR_i-r}  - [f(\bX_{i,r}, \epsilon_{i,\bar{r}, b'},r)]_{\bR_i-r}\right\|  
\Bigg\}
\Bigg].
\label{eq::emploss_MCAR}
\end{aligned}
\end{equation}
This construction allows partially observed units to contribute to the training. Its validity relies on MCAR and the strict propriety of the energy score, as formalized below.

\begin{theorem}
\label{thm::mcar}
Let $f^*$ be the population risk minimizer of the MCAR emputation risk \eqref{eq::emploss_MCAR}, \emph{i.e.},
$
f^* = \argmin_{f\in\mathcal{F}} \E[\hat{\mathcal{L}}(f)].
$
For any response pattern $r \in \mathcal{R}$ and a given $(x_r,r)$,
let 
$
\hat X_{\bar r} = [f^*(x_r, \epsilon_{\bar r}, r)]_{\bar r} 
$
be the output from $f^*$, where $\epsilon_{\bar r}$ is an independent standard Gaussian noise vector.
Under MCAR, the density of $\hat X_{\bar r}$ satisfies $p_{f^*}(x_{\bar r}|x_r,r) = p(x_{\bar r}|x_r)$.
\end{theorem}

\autoref{thm::mcar} shows that the population minimizer $f^*$ induced by \eqref{eq::emploss_MCAR} recovers the target extrapolation density under MCAR. By incorporating partially observed samples into training, the MCAR emputation risk exploits all available observed entries rather than discarding incomplete cases, improving efficiency without sacrificing imputation validity.

\subsection{Emputation under CCMV}
\label{sec::ccmv}
While MCAR provides a useful benchmark, it is often too restrictive for real-world datasets. In this section, we extend emputation to a popular MNAR scenario: the CCMV assumption \citep{little1993pattern}, also referred to as extended MAR (EMAR) in \citet{naf2024good}. Under CCMV, the extrapolation density is identified by the corresponding conditional distribution among complete cases:
\begin{equation}
p(x_{\bar r}|x_r,r) = p(x_{\bar r}|x_r,R = 1_d).
\label{eq::ccmv}
\end{equation} 

Since the extrapolation density for each incomplete pattern is identical to the density for complete cases, our selection function $S_r(\bR_i)$ needs to restrict training to observations with $\bR_i = 1_d$. 
For a target pattern $r$, the variables to be masked among complete cases are precisely those that would be missing under pattern $r$, namely variables in $\bar r$. Thus, under CCMV, we set the masked pattern $M_r(\bR_i) = 1_d-r = \bar r$ and the selection function $S_r(\bR_i) = I(\bR_i = 1_d) / |\bar r|$. Here, the factor $|\bar r|$ normalizes by the number of masked variables, preventing patterns with larger $|\bar r|$ from dominating the risk merely because more variables are masked.

The resulting empirical objective function is 
\begin{equation}
\begin{aligned}
\hat{\mathcal{L}}(f)
&= \frac{1}{n}\sum_{i=1}^n\Bigg[
\sum_{r\in\mathcal{R}}I(\bR_i=1_d) \frac{1}{|\bar r|}
\Bigg\{
\frac{1}{B}\sum_{b=1}^B \left\|
\bX_{i,\bar r} - [f(\bX_{i,r}, \epsilon_{i,\bar{r}, b}, r)]_{\bar r}\right\| \\
&-\frac{1}{2B(B-1)}\sum_{b, b'=1}^B \left\|[f(\bX_{i,r}, \epsilon_{i,\bar{r}, b}, r)]_{\bar r}  - [f(\bX_{i,r}, \epsilon_{i,\bar{r}, b'},r)]_{\bar r}\right\| 
\Bigg\}
\Bigg].
\label{eq::emploss_ccmv}
\end{aligned}
\end{equation}

The following theorem establishes its population target.

\begin{theorem}
\label{thm::ccmv}  
Let $\hat{\mathcal{L}}(f)$ be defined as in \eqref{eq::emploss_ccmv}, and let $f^*$ be the population minimizer
$
f^* = {\sf argmin}_{f} \E[\hat{\mathcal{L}}(f)].
$
For any response pattern $r \in \mathcal{R}$  and a given $(x_r,r)$,
let 
$
\hat X_{\bar r} = [f^*(x_r, \epsilon_{\bar r}, r)]_{\bar r} 
$
be the output from $f^*$, where $\epsilon_{\bar r}$ is an independent standard Gaussian noise vector.
Then the density of $\hat X_{\bar r}$ satisfies $p_{f^*}(x_{\bar r}|x_r,r) = p(x_{\bar r}|x_r, R=1_d)$.
\end{theorem}

\autoref{thm::ccmv} shows that the population minimizer of the CCMV emputation risk recovers the correct extrapolation density implied by CCMV. 
Compared to the MCAR construction in \autoref{thm::mcar}, the form of the emputation risk changes only through the masked pattern $M_r$ and the selection function $S_r$. This illustrates the central principle of emputation: different missing data assumptions can be implemented by modifying the masking patterns and selection functions while keeping the same underlying energy-score learning framework.

\subsection{Emputation for tree graphs}
\label{sec::tree}
The MCAR and CCMV examples already illustrate the key principle behind emputation: the masked pattern $M_r$ and selection function $S_r$ are determined by how a missing data assumption identifies the extrapolation density. Once its identifying assumption is specified in the pattern mixture framework, we can translate its identification strategy into the associated $M_r$ and $S_r$. 
To illustrate this idea beyond MCAR and CCMV, we consider the tree graph in \cite{suen2026modelingmultivariatemissingnesstree}, a special case of pattern graphs \citep{chen2022pattern}. 
A tree graph is a directed graph $G$ whose vertex set is $\mathcal{R}\cup\{1_d\}$. Each incomplete pattern $r\in \mathcal{R}$ has a unique parent ${\sf PA}(r)\in\mathcal{R}\cup\{1_d\}$
satisfying ${\sf PA}(r) > r$. The only exception is the vertex $1_d$, which has no parent and acts as the root of the graph $G$. 
In this paper, we treat the tree graph as given, and discussion of tree graph selection techniques can be found in \cite{suen2026modelingmultivariatemissingnesstree}.

Under a tree graph, each incomplete pattern $r$ borrows information from its unique parent through a pattern mixture model assumption
\begin{equation}
p(x_{\bar r}|x_r,R=r)=p(x_{\bar r}|x_r,R={\sf PA}(r)).
\label{eq::tree}
\end{equation}
CCMV is a special case of tree graphs where every incomplete vertex has the same parent ${\sf PA}(r) = 1_d$.
While \eqref{eq::tree} identifies the extrapolation density nonparametrically, the right-hand side may still involve variables that are unobserved under the parent pattern. It is thus unclear how to construct the masked pattern $M_r$ and selection function $S_r$ accordingly.
To resolve this problem, we consider a path decomposition implied by \eqref{eq::tree} to construct the emputation risk.

Under a tree graph $G$, every pattern $r\in\mathcal{R}$ is connected to the complete-case pattern by a unique path of arrows, denoted by $r=\gamma_0\leftarrow \gamma_1\leftarrow\dots\leftarrow \gamma_{K_r}=1_d$, where $\gamma_{k+1} = {\sf PA}(\gamma_k)$ for $k=0,\dots,K_r-1$.
For each directed arrow ${\sf PA}(\gamma)\to \gamma$, \eqref{eq::tree} implies 
\begin{equation}
p(x_{{\sf PA}(\gamma)-\gamma}|x_\gamma,R=\gamma)=p(x_{{\sf PA}(\gamma)-\gamma}|x_\gamma,R={\sf PA}(\gamma)),
\label{eq::tree2}
\end{equation}
where ${\sf PA}(\gamma)-\gamma$ indexes the coordinates that become observed when moving from $\gamma$
to its parent ${\sf PA}(\gamma)$. Clearly, each factor $p(x_{{\sf PA}(\gamma)-\gamma}|x_\gamma,R={\sf PA}(\gamma))$ is identifiable, provided $p(R={\sf PA}(\gamma))>0$. 
Equations \eqref{eq::tree} and \eqref{eq::tree2} imply that the extrapolation density factorizes along the unique path into
\begin{equation}
\begin{aligned}
 p(x_{\bar{r}} \mid x_r, r) &= \prod_{k=0}^{K_r-1} p(x_{{\sf PA}(\gamma_k) -\gamma_k} \mid x_{\gamma_k}, R = {\sf PA}(\gamma_k))\\
 &= \prod_{k=0}^{K_r-1} p(x_{\gamma_{k+1} -\gamma_k} \mid x_{\gamma_k}, R = \gamma_{k+1}).
 \end{aligned}
\label{eq::tree3}
\end{equation}
Therefore, it suffices to train our imputation model for the one-step conditional distributions under \eqref{eq::tree2}, and the full extrapolation distribution is then obtained by sequential imputation along the path in \eqref{eq::tree3}.

Equation \eqref{eq::tree2} suggests setting the masked pattern to be $M_r(\bR_i) = {\sf PA}(r) - r$
and the selection function to $S_r(\bR_i) = \frac{1}{|{\sf PA}(r)-r|}I(\bR_i ={\sf PA}(r))$. This choice ensures that only observations with response pattern $\bR_i = {\sf PA}(r)$ are used in training, and the masked coordinates are those observed in the parent pattern but not in pattern $r$.
Then, the emputation risk under a tree graph $G$ is
\begin{equation*}
\begin{aligned}
\hat{L}_G(f) &= \frac{1}{n} \sum_{i=1}^n \sum_{r\in\mathcal{R}}
\frac{I(\bR_i = {\sf PA}(r))}{|{\sf PA}(r)-r|}
\Bigg\{
    \frac{1}{B} \sum_{b=1}^B \left\| \bX_{i, {\sf PA}(r)-r} - 
    [f(\bX_{i,r}, \epsilon_{i,\bar{r},b}, r)]_{{\sf PA}(r)-r} \right\|\\
    &\quad    - \frac{1}{2B(B-1)} \sum_{b \neq b'} \left\| 
    [f(\bX_{i,r}, \epsilon_{i,\bar{r},b}, r)]_{{\sf PA}(r)-r}
    - [f(\bX_{i,r}, \epsilon_{i,\bar{r},b'}, r)]_{{\sf PA}(r)-r} \right\|
\Bigg\}.
\end{aligned}
\end{equation*}

\begin{theorem}
Let $f^* = \argmin_f \mathbb{E}[\hat{L}_G(f)]$ be the population risk minimizer of the tree graph $G$.
For any response pattern $r \in \mathcal{R}$  and a given $(x_r,r)$,
let 
$
\hat X_{{\sf PA}(r) -r} = [f^*(x_r, \epsilon_{\bar r}, r)]_{{\sf PA}(r) -r} 
$
be the output from $f^*$, where $\epsilon_{\bar r}$ is an independent standard Gaussian noise vector.
Then the density of $\hat X_{{\sf PA}(r) -r}$ satisfies
 $p_{f^*}(x_{{\sf PA}(r) -r}|x_r, R=r) = p(x_{{\sf PA}(r) -r}|x_r, R={\sf PA}(r))$.
\label{thm::treegraph}
\end{theorem}

Under the tree graph, every pattern $r \in \mathcal{R}$ 
has a unique path $r = \gamma_0 \leftarrow \gamma_1 \leftarrow \dots \leftarrow \gamma_{K_r} = 1_d$.
\autoref{thm::treegraph} implies that the extrapolation density
\begin{align*}
    p_{f^*}(x_{\bar{r}} \mid x_r, r) &:= \prod_{k=0}^{K_r-1} 
    p_{f^*}(x_{{\sf PA}(\gamma_k) - \gamma_k} \mid x_{\gamma_k}, R=\gamma_k) \\
    &=\prod_{k=0}^{K_r-1} 
    p(x_{{\sf PA}(\gamma_k) - \gamma_k} \mid x_{\gamma_k}, R={\sf PA}(\gamma_k))\\ 
    & \overset{\eqref{eq::tree3}}{=}p(x_{\bar{r}} \mid x_r, r)
\end{align*}
recovers the true extrapolation density $p(x_{\bar{r}} \mid x_r, r)$ for all $r\in\mathcal{R}$.

After training $\hat f$, an incomplete observation $(X_R,R)$ is imputed by iterating the following three steps until no missing entries are left:
\begin{enumerate}
\item Set $\gamma = R$ and find its parent ${\sf PA}(\gamma)$ in the tree graph. 
\item Impute the missing entries $X_{{\sf PA}(\gamma)-\gamma}$ via $\hat X_{{\sf PA}(\gamma)-\gamma} = [\hat f(X_\gamma, \epsilon_{\bar{\gamma}}, \gamma)]_{{\sf PA}(\gamma) -\gamma}$, where $\epsilon_{\bar \gamma}$ is an independent Gaussian noise vector with each element drawn from $N(0,1)$.
\item Update $R = {\sf PA}(\gamma)$. Return to Step 1 if $R\neq 1_d$. 
\end{enumerate}
The above procedure is a sequential imputation scheme following the product formula in \eqref{eq::tree3}.

\section{Emputation for Monotone Missingness}
\label{sec::monotone}

We next discuss how to implement emputation for monotone missingness under several conventional assumptions.
Monotone missingness arises naturally in longitudinal studies and clinical trials with dropout, where once a variable is missing, all subsequent variables are also missing \citep{daniels2008missing}. Namely,
for any $r\in \mathcal{R}$, if $r_j = 0$, then $r_{k} = 0$
for all $k>j$. 
In this setting, each response pattern $r$ can be summarized by its dropout time $|r|$, so the observed data can be written as $(X_{\leq T}, T)$, where $T=|R| \in \{1,2,...,d\}$ and $X_{\leq T} = (X_1,\dots, X_T)$. For a sample of size $n$, the observed data are thereby
$
(\bX_{1, \leq\bT_1}, \bT_1),\cdots, (\bX_{n, \leq \bT_n}, \bT_n),
$
where $\bT_i = |\bR_i|$.
For a unit with dropout time $T=t$, the extrapolation density is
$
p(x_{>t}|x_{\leq t}, t)
$,
which admits the sequential factorization
\begin{equation}
p(x_{>t}|x_{\leq t}, t) = \prod_{s=t}^{d-1} p(x_{s+1}|x_{\leq s}, t).
\label{eq::mono}
\end{equation}
This factorization is analogous to the tree graph's factorization in \eqref{eq::tree3}.
Thus, identifying the extrapolation density reduces to identifying each component $p(x_{s+1}|x_{\leq s},t)$ for $s\geq t$. Different identifying assumptions correspond to different choices of the reference distribution \citep{daniels2023bayesian}. To emphasize that these assumptions are specialized to monotone missingness, we use the prefix ``m-''. We focus on three commonly used assumptions. 
\begin{itemize}
\item {\bf Complete-case missing value (m-CCMV).} 
Similar to CCMV, the m-CCMV assumption identifies each component using complete cases:
$$
p(x_{s+1}|x_{\leq s},T=t) = p(x_{s+1}|x_{\leq s},T=d). 
$$

\item {\bf Available-case missing value (m-ACMV).}
Since m-CCMV relies only on complete cases, it discards observations in which $X_{s+1}$ is observed but later variables are missing. The m-ACMV assumption instead uses all available cases in which $X_{s+1}$ is observed:
\begin{equation}
p(x_{s+1}|x_{\leq s},T=t) = p(x_{s+1}|x_{\leq s},T\geq s+1).
\label{eq::acmv}
\end{equation}
\cite{molenberghs1998missing} showed that m-ACMV is equivalent to MAR under monotone missingness.

\item {\bf Nearest-case missing value (m-NCMV).}
The m-NCMV assumption identifies each component using the nearest response pattern, namely the pattern with exactly one additional observed variable:
$$
p(x_{s+1}|x_{\leq s},T=t) = p(x_{s+1}|x_{\leq s},T=s+1). 
$$
\end{itemize}
Under all three assumptions,
the right-hand side is an identifiable conditional distribution of the form $p(x_{s+1}|x_{\leq s}, T= \omega(s))$ for some reference set $\omega(s)\subset\{s+1,s+2,\cdots, d\}$ that depends on $s$ but not on the original dropout time $t$. 
Therefore, under these assumptions, the imputation model only needs to learn the conditional distribution $p(x_{s+1}|x_{\leq s}, T= \omega(s))$ associated with each $s=1,\cdots, d-1$.

\subsection{Emputation under m-ACMV} 
\label{sec::monotone_acmv}
We first show how the m-ACMV restriction can be incorporated into the emputation framework. This case is of particular interest because m-ACMV is equivalent to MAR under monotone missingness \citep{molenberghs1998missing}. The m-CCMV and m-NCMV cases can be derived in a similar manner.

The m-ACMV assumption in \eqref{eq::acmv} identifies the one-step extrapolation density as $p(x_{s+1}|x_{\leq s},T=t) = p(x_{s+1}|x_{\leq s},T\geq s+1)$
using observations with a dropout time $T\geq s+1$. 
Since the target is a conditional density of variable $X_{s+1}$, our masked pattern is $M_r(\bR_i) \equiv M_s(\bT_i) = s+1 \equiv |r|+1$, and 
the selection function $S_r(\bR_i) \equiv S_s(\bT_i) = I(\bT_i \geq s+1) \equiv I(|\bR_i| \geq |r|+1)$.
For monotone missingness, we write the masked coordinate as $M_s(\bT_i) \in \{2,\cdots, d\}$ for $s=1,\cdots, d-1$ 
and the selection function as $S_s(\bT_i)$.

While one could use the same vector-valued neural network imputation model $f$ as in the nonmonotone setting of Section~\ref{sec:nonmon_emputation}, monotone missingness allows a simpler parameterization.
Since monotone response patterns are indexed by the dropout time $s =|r|$, we use $s$ directly as an input coordinate to the imputation model. Specifically, let $g: \R^d\times \{1,2,\cdots, d\}\to \R$ be a neural network with a scalar output. Sampling from the conditional distribution $p_g(x_{s+1}|x_{\leq s}, s)$ is done by evaluating $g(x_{\leq s}, \epsilon_{>s}, s)$, where $\epsilon_{>s}$ is an independent standard Gaussian noise vector.

The empirical emputation risk under m-ACMV is
\begin{equation}
\begin{aligned}
\hat{\mathcal{L}}(g)&= \frac{1}{n}\sum_{i=1}^n\sum_{s=1}^{d-1}
 S_s(\bT_i)
 \Bigg\{ 
 \frac{1}{B}\sum_{b=1}^B \left|
 \bX_{i,M_{s}(\bT_i)}  - g(\bX_{i,\leq s}, \epsilon_{i, > s, b}, s)\right| \\
 & \quad -\frac{1}{2B(B-1)}\sum_{b\neq b'} \left|g(\bX_{i,\leq s}, \epsilon_{i, > s, b}, s)  - g(\bX_{i,\leq s}, \epsilon_{i, > s, b'}, s)\right| \Bigg\}\\
 &= \frac{1}{n}\sum_{i=1}^n\sum_{s=1}^{d-1}
 I(\bT_i >s)
 \Bigg\{ 
 \frac{1}{B}\sum_{b=1}^B \left|
 \bX_{i,s+1}  - g(\bX_{i,\leq s}, \epsilon_{i, > s, b}, s)\right| \\
 & \quad -\frac{1}{2B(B-1)}\sum_{b\neq b'} \left|g(\bX_{i,\leq s}, \epsilon_{i, > s, b}, s)  - g(\bX_{i,\leq s}, \epsilon_{i, > s, b'}, s)\right| \Bigg\}
\label{eq::mono::ER1}
\end{aligned}
\end{equation}
where the noise vectors $\{\epsilon_{i, >s, b}\}_{b=1}^B$ are independent draws from a standard Gaussian distribution $N(0,\mathbf{I}_{d-s})$.

\begin{theorem}
Let $g^* = \argmin_{g} \E[\hat{\mathcal{L}}(g)]$ be the population risk minimizer of the emputation risk. 
For any given $(x_r,r) \equiv (x_{\leq s}, s)$ with $s=|r|$, 
let 
$
\hat X_{s+1} = [g^*(x_{\leq s}, \epsilon_{>s}, s)]
$
be the output of $g^*$, where $\epsilon_{>s}$ is an independent standard Gaussian noise vector. 
Then, the density of $\hat X_{s+1}$ satisfies 
$p_{g^*}(x_{s+1}|x_{\leq s}, T = s) = p(x_{s+1}|x_{\leq s}, T \geq s+1)$, which is the target extrapolation density under m-ACMV.
\label{thm::monotone::acmv}
\end{theorem}

Because of the sequential factorization in \eqref{eq::mono}, learning the one-step conditional distribution of \(X_{s+1}\) given \((x_{\leq s},s)\) determines the full extrapolation distribution. 
Thus, \autoref{thm::monotone::acmv} establishes the valid population target of 
emputation under m-ACMV.
After training $\hat g$, an incomplete observation $(X_{\leq T},T)$, where $T=|R|$, is imputed sequentially as in the tree graph procedure.
Specifically, starting from $s=T$, we iterate the following two steps until no missing entries are left:
\begin{enumerate}
\item Impute the missing entry $X_{s+1}$ by $\hat X_{s+1} = \hat g(X_{\leq s}, \epsilon_{>s}, s)$, where $\epsilon_{>s}$ is an independent standard Gaussian noise vector.
\item Update $s \gets s+1$.  If $s < d$, return to Step 1. 
\end{enumerate}
This sequential procedure follows the factorization in \eqref{eq::mono}.
Because m-ACMV is equivalent to MAR \citep{molenberghs1998missing} under monotone missing data, this construction under emputation offers a flexible neural network imputation model under MAR.

\subsection{Emputation under m-CCMV and m-NCMV}
\label{sec::monotone_ccmv_ncmv}
The same construction applies to m-CCMV and m-NCMV by changing only the selection rule in the emputation risk. In all three monotone assumptions considered here, the masked coordinate remains $X_{s+1}$.
Thus, the masked coordinate is still $M_s(\bT_i)=s+1$, while the selection function changes according to the identifying assumption.
Under m-CCMV, each extrapolation density is identified using complete cases, so the selection function is $S_s(\bT_i) = I(\bR_i = 1_d) = I(\bT_i  = d)$. 
Under m-NCMV, each extrapolation density is identified using the nearest available pattern, namely observations with exactly one additional observed variable. Hence, the selection function is $S_s(\bT_i) =  I(\bT_i  = s+1)$.

\begin{corollary}
\label{cor::monotone_ccmv_ncmv}
Let $g^*$ be the population minimizer of the emputation risk under m-CCMV or m-NCMV. 
For any given $(x_r,r) \equiv (x_{\leq s}, s)$ with $s=|r|$, 
let 
$
\hat X_{s+1} = [g^*(x_{\leq s}, \epsilon_{>s}, s)]
$
be the output of $g^*$, where $\epsilon_{>s}$ is an independent standard Gaussian noise vector. 
Then, the density of $\hat X_{s+1}$ satisfies 
\[
p_{g^*}(x_{s+1}\mid x_{\leq s},s)
=
\begin{cases}
p(x_{s+1}\mid x_{\leq s},T=d), & \text{under the m-CCMV risk},\\
p(x_{s+1}\mid x_{\leq s},T=s+1), & \text{under the m-NCMV risk}.
\end{cases}
\]
\end{corollary}

Corollary \ref{cor::monotone_ccmv_ncmv} shows that 
the emputation framework targets the appropriate extrapolation density under both m-CCMV and m-NCMV, together with the sequential factorization in \eqref{eq::mono}. Thus, as in the m-ACMV case, emputation accommodates different monotone missing-data assumptions by modifying the selection function while keeping the same masking structure and energy-score learning framework.

\section{Selecting a Missing Data Assumption}
The preceding sections show that the construction of the emputation risk is guided by the chosen missing data assumption. 
We now briefly discuss how such a missing data assumption may be selected in practice.
Except for MCAR, the missing data assumptions considered in this paper, including CCMV, pattern graphs, and the monotone missingness assumptions considered above, are nonparametric identifying assumptions \citep{robins1997MNAR,vansteelandt2006ignorance,daniels2008missing}. As such, they do not impose testable restrictions on the observed-data distribution and therefore cannot be rejected using the observed data alone.
This is a desirable property in the sense that these assumptions are weak and always compatible with the observed data. 
However, it is also a limitation that we cannot purely use any data-driven criterion to select an assumption. Therefore, the choice of a missing data assumption relies primarily on our domain knowledge of the data-generating process and the missingness mechanism. 

For instance, although we introduce CCMV through the pattern mixture model formulation, it also admits a selection model representation and can be interpreted as a logit discrete choice model, as explained in \cite{discreteEric2018}. Thus, CCMV may be appropriate when complete cases provide a scientifically plausible reference distribution for incomplete cases. 
Tree graphs \citep{suen2026modelingmultivariatemissingnesstree} also admit a selection odds formulation and may be useful when the missingness mechanism is better represented by borrowing information along a structured hierarchy of response patterns rather than directly from complete cases.

For monotone missingness, the choice among m-ACMV, m-CCMV, and m-NCMV depends on which observed patterns are considered plausible references for the missing values. The m-ACMV assumption is equivalent to MAR under monotone missingness \citep{molenberghs1998missing, molenberghs2007missing}.
Therefore, when MAR is believed to be the actual missing mechanism, m-ACMV provides the corresponding emputation risk. 
The m-NCMV assumption has a nearest-donor interpretation, since it identifies each missing component using the nearest response pattern \citep{daniels2023bayesian}. This assumption may be justified when individuals who drop out at adjacent times are expected to be similar, so that subjects dropping out at times $t$ and $t+1$ provide the most relevant donor information.

In practice, we recommend using substantive knowledge to identify a small set of plausible assumptions and then assessing the sensitivity of downstream conclusions across the corresponding emputation risks. We illustrate this strategy in the real-data application in Section~\ref{sec::realdata}.

\section{Simulation Studies}
\label{sec::experiments}
We conduct simulation studies to evaluate the performance of emputation under different missing data assumptions. We compare it with mean imputation, Expectation--Maximization (EM) with a multivariate Gaussian model, multiple imputation by chained equations (MICE; \citealt{raghunathan2001multivariate, van1999flexible}), missForest (MF; \citealt{stekhoven2012missforest}), and the generative adversarial imputation network (GAIN; \citealt{yoon2018gain}). Among deep generative imputation methods, we include GAIN as the main baseline because it is a representative deep generative approach for imputation. The experiments are conducted on three benchmark datasets from the UCI Machine Learning Repository: Concrete, Wine, and CCPP. These datasets vary in sample size and dimension, as summarized in \autoref{tab::sim_data}. 
\begin{table}[t]
\centering
\caption{Datasets used in the simulation, with sample size $n$ and number of variables $d$.}
\label{tab::sim_data}
\small
\setlength{\tabcolsep}{5pt}
\renewcommand{\arraystretch}{0.9}
\begin{tabular}{@{}lrr@{}}
\toprule
Dataset & \multicolumn{1}{c}{$n$} & \multicolumn{1}{c}{$d$} \\
\midrule
Concrete & 1030 & 9  \\
Wine     & 4898 & 12 \\
CCPP     & 9568 & 5  \\
\bottomrule
\end{tabular}
\end{table}

For each dataset, missing values are generated under MCAR and CCMV with an overall missingness rate of approximately 20\%. Under MCAR, entries are independently set to missing. Under CCMV, response patterns are generated using the selection model described in the Supplementary Material. MCAR is included as a baseline mechanism because it is widely used in imputation simulation studies. We also include CCMV because it represents an MNAR setting aligned with the emputation framework. We focus on these two assumptions because other missing data restrictions are less straightforward to benchmark. For example, tree graph assumptions require specifying a graph structure over response patterns, whereas monotone restrictions are designed for monotone dropout settings and are less suitable for the general nonmonotone missingness considered here. 
Additional details on the missingness generation process, along with results under MAR, 40\% missingness, and varying complete-case proportions, are reported in the Supplementary Material.

Results are averaged over 100 independent repetitions. For multiple imputation methods, we generate 10 imputed datasets.
The MICE benchmark is implemented using the default methods in the R package \texttt{mice}, which uses predictive mean matching for continuous variables. The emputation model is implemented as a feedforward neural network with $L=3$ hidden layers, each of width 500, and is trained for 500 epochs using a learning rate of $10^{-4}$. We set the number of Monte Carlo samples used to approximate the energy score to $B=2$, which we found to work well in practice.
GAIN is trained for 1{,}000 iterations with a batch size of 128, a hint rate of 0.9, and $\alpha = 10$.

We evaluate imputation performance using both pointwise and distributional criteria.
To assess the preservation of dependence and distributional agreement, we report the mean absolute difference in correlations (MADC), energy distance, and squared maximum mean discrepancy (MMD$^2$). Details of the evaluation metrics and pointwise accuracy results are given in the Supplementary Material. 
The implementation code is available at \url{https://github.com/yjyang00/emputation}.

\begin{figure}[ht]
    \centering
    \includegraphics[width=\linewidth]{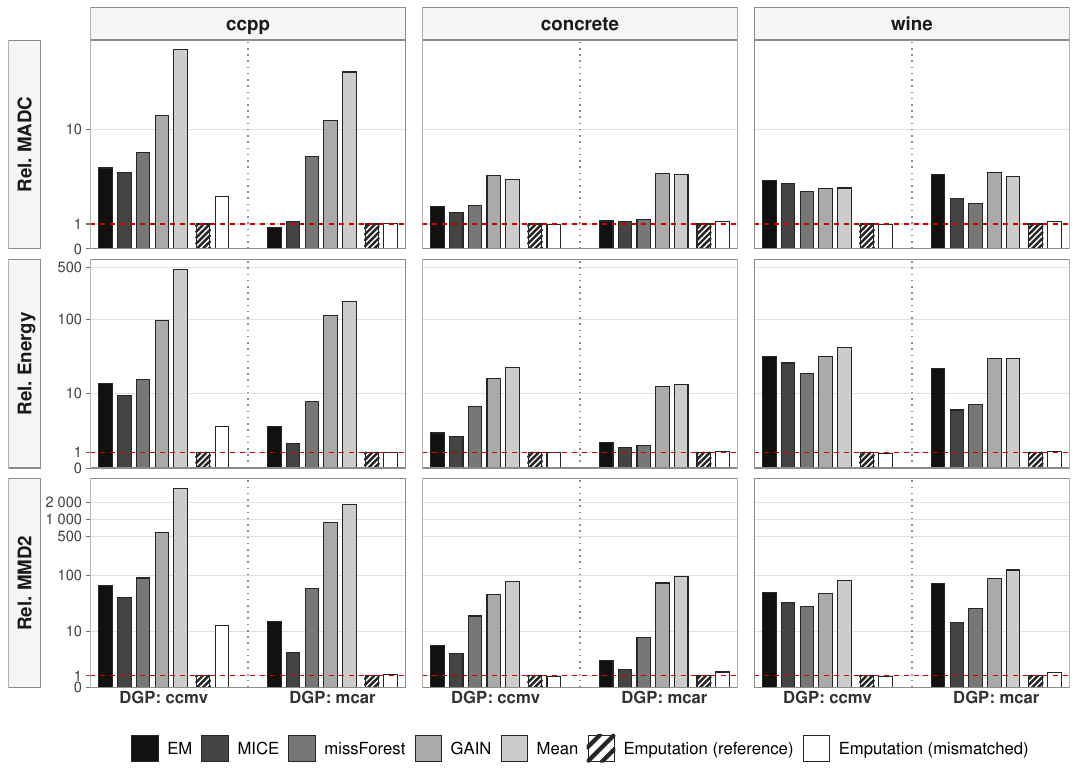}
    \caption{Simulation under 20\% missingness: performance of imputation methods in MADC, energy distance (relative to emputation), and $\text{MMD}^2$ (relative to emputation). The y-axis is shown on a logarithmic scale to accommodate the large variation across methods. The x-axis shows the missingness-generation process.}
    \label{fig:sim}
\end{figure}

\paragraph*{Results.}
\autoref{fig:sim} reports the main simulation results under 20\% missingness. Emputation achieves the strongest distributional performance in most settings. When the training risk matches the true missing data generation process, emputation attains the smallest energy distance and MMD$^2$ across all three datasets, as well as the smallest MADC in most cases. In contrast, mean imputation and GAIN often distort the target distribution, while MICE and missForest generally exhibit larger distributional discrepancies than emputation.

The comparison between matched and mismatched emputation risks illustrates the role of the missing data assumption. As expected, the best performance is typically obtained when the training risk is aligned with the data-generating mechanism. Nevertheless, the mismatched emputation risk remains competitive in many cases, suggesting that the energy-score-based objective is reasonably robust to moderate misspecification of the working missing data assumption.

\section{Real Data Application}
\label{sec::realdata}
We apply emputation to data from the National Alzheimer’s Coordinating Center (NACC). We use the Uniform Data Set (UDS), Neuropathology (NP) Data Set, and APOE genotype data, all publicly available upon request at \url{https://naccdata.org}. Our goal is to investigate predictors of cognitive resilience and resistance in the presence of Alzheimer's disease (AD) neuropathology. Cognitive resilience refers to an individual's ability to preserve cognition despite significant AD pathology, while cognitive resistance refers to an individual's ability to maintain cognitive normality in the presence of low AD pathology. Understanding the demographic, genetic, and clinical correlates of these phenotypes may illuminate protective mechanisms against dementia.

The analytic sample is restricted to participants with brain autopsy data (\texttt{NACCAUTP}=1). To align clinical assessments with end-of-life cognitive status, only visits occurring within two years prior to death are retained, and the visit closest to death is selected for each participant. Death dates are approximated as the last day of the recorded month. Participants with unknown APOE genotype (\texttt{NACCAPOE}=9) are excluded, as genotype is a fixed genetic attribute that is not reliably predictable from observed clinical or demographic variables. 

AD neuropathology is measured using the Alzheimer's Disease Neuropathologic Change (ADNC) score, a four-level ordinal scale (0=not AD, 1=low, 2=intermediate, 3=high). 
For subjects with missing ADNC, the score is derived from Braak stage and CERAD neuritic plaque score following the semi-quantitative classification of \cite{brenowitz2017alzheimer}. 
We consider demographic (age at death, sex, race, education), genetic (APOE $\varepsilon$4 and $\varepsilon$2 carrier status), and clinical variables (body mass index (BMI), systolic blood pressure, geriatric depression scale (GDS), and diuretic use) as predictors \citep{walker2023cognitive, sin2023characteristics}. The final analytic sample consists of 4{,}906 subjects, with characteristics across complete cases and incomplete cases summarized in \autoref{tab:cc_EDA}.

A central challenge is missingness in clinical covariates. Body mass index, systolic blood pressure, and geriatric depression scale have 40\% to 50\% missingness, and only approximately 36\% of subjects are observed completely across these measures. To address missingness, we apply emputation under MCAR and CCMV and compare it with MICE and complete-case analysis. MICE is implemented using the default predictive mean matching method for variables subject to missingness.

\begin{table}[t]
\centering
\caption{Characteristics by complete and incomplete case status. Age, education, and CDR Global Score are summarized by mean (SD); other variables are summarized by count (\%).}
\label{tab:cc_EDA}
\footnotesize
\setlength{\tabcolsep}{4pt}
\renewcommand{\arraystretch}{0.9}
\begin{tabular}{lcc}
\toprule
 & \multicolumn{1}{c}{Complete} & \multicolumn{1}{c}{Incomplete} \\
 & \multicolumn{1}{c}{$N=1820$} & \multicolumn{1}{c}{$N=3086$} \\
\midrule
Age at death, y & 82.86 (10.32) & 79.76 (11.71) \\
Female, n (\%) & 767 (42.1) & 1459 (47.3) \\
White, n (\%) & 1720 (94.5) & 2921 (94.7) \\
Education, y & 15.51 (3.02) & 15.44 (3.15) \\
APOE $\varepsilon$4 carrier, n (\%) & 674 (37.0) & 1401 (45.4) \\
APOE $\varepsilon$2 carrier, n (\%) & 241 (13.2) & 330 (10.7) \\
ADNC, n (\%) & & \\
\quad Not AD & 188 (10.3) & 334 (10.8) \\
\quad Low & 715 (39.3) & 714 (23.1) \\
\quad Intermediate & 401 (22.0) & 511 (16.6) \\
\quad High & 516 (28.4) & 1527 (49.5) \\
CDR Global Score & 0.97 (0.88) & 2.23 (1.02) \\
\bottomrule
\end{tabular}
\end{table}

\paragraph*{Analysis model.}
We fit separate logistic regression models for cognitive resilience and resistance.
The resilience model is restricted to subjects with intermediate or high ADNC (\texttt{NPADNC}$\in\{2,3\}$), where resilience is defined as cognitive normality, measured by a Clinical Dementia Rating (CDR) global score of zero (\texttt{CDRGLOB}$=0$), and non-resilience is defined as cognitive impairment (\texttt{CDRGLOB}$\geq 0.5$).
The resistance model is restricted to those with low or absent ADNC (\texttt{NPADNC}$\in\{0,1\}$), with resistance defined as cognitive normality (\texttt{CDRGLOB}=0) and non-resistance defined as cognitive impairment (\texttt{CDRGLOB} $\geq 0.5$). For complete-case analysis (CC), the logistic regression is fitted on the complete cases within each of the $B=500$ bootstrap replicates. For emputation, the bounded variables subject to missingness, including body mass index, systolic blood pressure, and geriatric depression scale, 
are first mapped to the real line using a logit transformation before imputation and then transformed back to their original scales using the inverse logit transformation after imputation. The emputation model is trained for 1000 epochs using a neural network with three hidden layers, each of width 500, and a learning rate of $10^{-4}$.
For MICE and emputation, 10 imputed datasets are drawn per bootstrap replicate and the logistic regression is fitted on each imputed dataset. Odds ratios (OR) and 95\% bootstrap confidence intervals are reported for all three methods.

\begin{figure}[t]
    \centering
    \includegraphics[width=\linewidth]{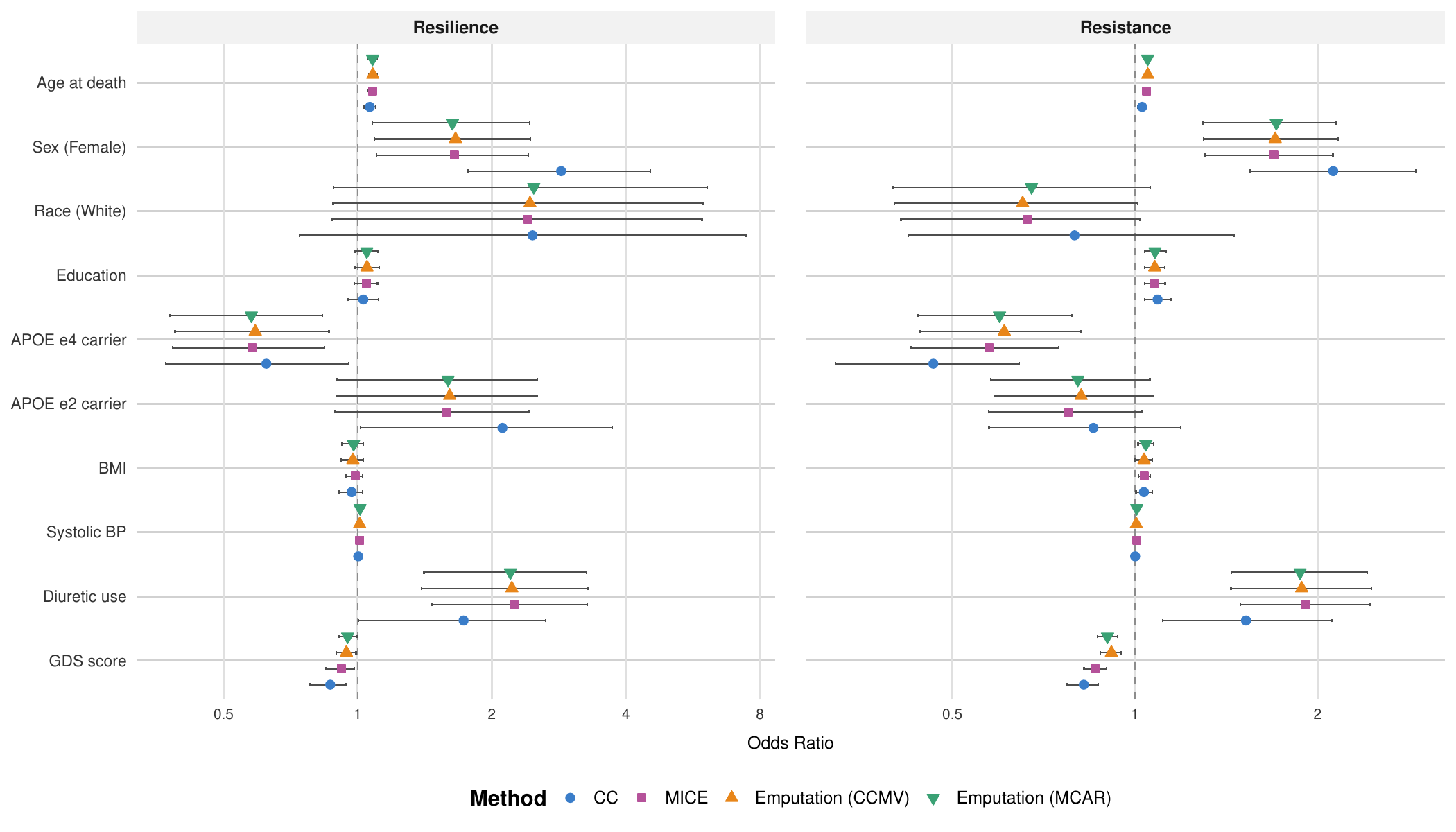}
    \caption{Odds ratios (ORs) and 95\% confidence intervals for predictors of cognitive resilience (left) and cognitive resistance (right) under four missing-data analyses: complete-case analysis (CC), multiple imputation by chained equations (MICE), emputation under MCAR, and emputation under CCMV.}
    \label{fig::nacc_forestplot}
\end{figure}

\paragraph*{Results.} \autoref{fig::nacc_forestplot} displays estimated odds ratios and confidence intervals for the resilience and resistance models.
In the resilience model, APOE $\varepsilon$4 carrier status is consistently associated with lower odds of cognitive resilience across all methods, in agreement with its established role as a genetic risk factor for AD \citep{farrer1997effects, belloy2023apoe}.
Female sex is associated with higher odds of resilience. The complete-case estimate is larger than those from MICE and emputation, which are closely aligned with each other. This discrepancy is consistent with the descriptive summary in \autoref{tab:cc_EDA} and may reflect a selection bias toward a ``healthier'' complete-case subsample, where complete cases are less cognitively impaired on average (mean CDR Global Score 0.97 vs. 2.23) and less likely to have high ADNC pathology (28.4\% vs.\ 49.5\%) compared to incomplete cases. Diuretic use shows positive associations with resilience under MICE and emputation, while the complete-case estimate is in the same direction but not statistically significant.

In the resistance model, APOE $\varepsilon$4 carrier status is associated with lower odds of resistance. Higher education is associated with higher odds of resistance, in line with the cognitive reserve hypothesis \citep{stern2012cognitive}.
Diuretic use shows a positive association under MICE and both emputation specifications, which agrees with prior evidence from \cite{sin2023characteristics}. GDS score is associated with lower odds of resistance under MICE and emputation, with complete-case analysis showing the same direction but a wider confidence interval. BMI shows a weak positive association, with confidence intervals remaining close to one, suggesting limited evidence for a strong association.

Overall, the main scientific findings are consistent across imputation methods and emputation specifications. In contrast, the discrepancies between imputation-based analyses and complete-case analysis highlight the potential impact of discarding incomplete observations. Additional sensitivity analyses and data descriptions are provided in the Supplementary Material.

\section{Conclusion} \label{sec::conclusion}
In this paper, we propose the emputation framework, which enables the training of a deep neural network model that recovers the correct imputation model under a missing data assumption. This framework offers insight into how to properly guide the training process of a neural network model using an identification assumption. Simulation studies demonstrate strong distributional performance relative to existing methods, and the Alzheimer's disease application illustrates its use in a real-world setting.

Several directions remain open for future work. First, the current framework covers MAR under monotone missingness through m-ACMV, but its connection to nonmonotone MAR remains unclear. Extending emputation to nonmonotone MAR as well as to other nonparametric identification assumptions would broaden the class of missing data assumptions that the framework can accommodate. Second, while the energy score provides a strictly proper scoring rule for distributional learning, other distributional objectives, such as other kernel scoring rules, could also be incorporated into the emputation framework.
Understanding how the choice of scoring rule affects imputation performance and theoretical guarantees is a natural direction for further study.
Third, it would be valuable to develop neural network models for selection models. The emputation framework can be viewed as a neural network model for pattern mixture identification, whereas selection models provide another common approach for MNAR problems and often offer more interpretable mechanisms \citep{diggle1994informative,ipsen2021notmiwae,daniels2023bayesian}. 
How to train neural generative imputation models under assumptions formulated through selection models remains an open problem.

{\small
\renewcommand{\baselinestretch}{1.5}\selectfont
\bibliography{ref}
}

\clearpage

\begin{center}
	{\Large \bf  Supplementary Materials for ``Emputation: Identification-Guided Neural Imputation Framework''}\\
\end{center}

\bigskip

\noindent

\appendix

\renewcommand{\thesection}{Online Appendix \Alph{section}}
\renewcommand{\thesubsection}{\Alph{section}.\arabic{subsection}}

\renewcommand{\thefigure}{S\arabic{figure}}
\renewcommand{\thetable}{S\arabic{table}}
\setcounter{figure}{0}
\setcounter{table}{0}

\appendix

\section{Additional Simulation Details}
\label{app:sim_details}
This section provides additional details on the simulation studies reported in the main paper. 

We generate missingness under MCAR, CCMV, and MAR. For each mechanism, we consider two overall missingness levels, 20\% and 40\%. 

For the MCAR setting, we first randomly select a proportion $c_c=20\%$ of rows to remain fully observed. 
Among the remaining proportion $1-c_c=0.8$ of rows, each entry is independently set to missing with probability $\rho$. 
Thus, the overall missing rate is approximately $(1-c_c)\rho$. 
We set $\rho=0.25$ for the 20\% missingness setting and $\rho=0.50$ for the 40\% missingness setting.

For the CCMV setting, missingness is generated using a pattern graph \citep{chen2022pattern}.
The response pattern set consists of the complete-case pattern and a collection of pre-specified incomplete patterns. Under a pattern graph, the selection odds relative to the complete-case pattern are
\[
\frac{p(R=r| X)}{p(R=1_d| X)}
=\frac{p(R=r| X_r)}{p(R=1_d| X_r)}=O_r(X_r)=
\exp(
\alpha_r + \beta_r^T X_r)
\]
and the pattern-specific probabilities are
\[
p(R=1_d| X)
=
\left[
1+\sum_{r\neq 1_d}
O_r(X_r)
\right]^{-1}, \quad p(R=r| X)=p(R=1_d| X)O_r(X_r).
\]
Each row is then assigned a response pattern according to these probabilities. We set $\alpha_r=-4.0$ for all $r$ and each component of $\beta_r$ equal to 1 to obtain approximately 20\% missingness. For the 40\% missingness setting, we set $\alpha_r=-0.5$ and each component of $\beta_r$ equal to 1.

For the MAR setting, we use a sequential construction adapted from the MAR simulation in \citet{yoon2018gain}. 
We first randomly select a proportion $c_c=0.2$ of rows to remain fully observed. 
Missingness is then introduced among the remaining rows. Let $N_m=(1-c_c)N$ denote the number of rows subject to missingness, and define the adjusted missingness rate as $\widetilde p^{\,m}=\frac{p^m}{1-c_c}$,
where $p^m=0.20$ or $0.40$ is the overall missingness rate.
The response pattern $\bR_i=(\bR_{i1},\ldots,\bR_{id})$ is generated sequentially for each row $i=1,\ldots,N_m$. 
For the first variable, we set $p(\bR_{i1}=0)=\widetilde p^{\,m}$.
For variable $j>1$, the probability that $\bR_{ij}=0$ depends on the previously generated response indicators and their corresponding observed values:
\[
p(\bR_{ij}=0)
=
\frac{
	\widetilde p^{\,m} N_m
	\exp\left[-\sum_{k<j}\left\{w_k \bR_{ik}X_{ik}
	+
	b_k(1-\bR_{ik})\right\}\right]
}{
	\sum_{\ell=1}^{N_m}
	\exp\left[-\sum_{k<j}\left\{w_k \bR_{\ell k}X_{\ell k}
	+
	b_k(1-\bR_{\ell k})\right\}\right]
},
\]
where $w_k,b_k\sim \mathrm{Unif}(0,1)$. This construction thereby induces MAR missingness.

We evaluate imputation performance from both pointwise and distributional perspectives. 
Let $\mathcal{P}$ denote the set of missing entries, let $S=100$ be the number of Monte Carlo repetitions, and let $K=10$ be the number of imputations per repetition. 
For repetition $s$ and imputation $k$, let $X^{\mathrm{imp}}_{i,j,s,k}$ denote the imputed value of entry $(i,j)$, and let $X^{\mathrm{true}}_{i,j}$ denote the corresponding true value. 

The root mean squared error (RMSE) is defined as
\[
\mathrm{RMSE}
=
\frac{1}{S}\sum_{s=1}^S
\sqrt{
	\frac{1}{|\mathcal{P}|}
	\sum_{(i,j)\in\mathcal{P}}
	\left(
	X_{i,j}^{\mathrm{true}}
	-
	\frac{1}{K}\sum_{k=1}^K X_{i,j,s,k}^{\mathrm{imp}}
	\right)^2
}.
\]
The mean absolute error (MAE) is defined as
\[
\mathrm{MAE}
=
\frac{1}{S}\sum_{s=1}^S
\frac{1}{|\mathcal{P}|}
\sum_{(i,j)\in\mathcal{P}}
\left|
X_{i,j}^{\mathrm{true}}
-
\frac{1}{K}\sum_{k=1}^K X_{i,j,s,k}^{\mathrm{imp}}
\right|.
\]

The mean absolute difference in correlations (MADC) is defined as
\[
\mathrm{MADC}
=
\frac{1}{S}\sum_{s=1}^S
\sqrt{
\frac{2}{d(d-1)}
\sum_{i<j}
\left|
C_{i,j}^{\mathrm{true}}
-
\frac{1}{K}\sum_{k=1}^K C_{i,j,s,k}^{\mathrm{imp}}
\right|
}.
\]
Here, $C^{\mathrm{true}}$ is the correlation matrix of the true complete data, and $C^{\mathrm{imp}}_{s,k}$ is the correlation matrix of the $k$th imputed dataset in repetition $s$.

Distributional agreement is assessed using the energy distance between the standardized true data and each imputed dataset. For two distributions $P$ and $Q$, the energy distance \citep{szekely2013energy} is defined as
\[
\text{ED}_{\beta}(P,Q)=2\mathbb{E}\|X - Y\|^{\beta} - \mathbb{E}\|X - X'\|^{\beta}
- \mathbb{E}\|Y - Y'\|^{\beta}
\]
where $\beta\in (0,2)$, $X,X'$ are independent draws from $P$, and $Y,Y'$ are independent draws from $Q$. We use $\beta=1$ throughout, which corresponds to the standard energy distance. We can estimate it using a sample version through
\[
\widehat{ED}(P,Q) = \frac{2}{n_X \cdot n_Y}\sum_{i,j}\|X_i - Y_j\| - 
\frac{1}{n_X^2}\sum_{i,j}\|X_i - X_j\| - 
\frac{1}{n_Y^2}\sum_{i,j}\|Y_i - Y_j\|.
\]
We report the average estimated energy distance over $S$ simulations 
and $K$ multiple imputations,
\[
\overline{\varepsilon} = \frac{1}{S}\sum_{s=1}^{S}\frac{1}{K}
\sum_{k=1}^{K}\varepsilon_{s,k},
\]
where $\varepsilon_{s,k}$ denotes the estimated energy distance between 
the standardized true data and the $k$-th imputed dataset in the $s$-th 
simulation. Each $\varepsilon_{s,k}$ is computed using the 
\texttt{eqdist.e} function from the \texttt{energy} package 
\citep{rizzo2022energy}, which returns $\frac{n}{2}\widehat{\text{ED}}$ 
when both samples have equal size $n$, where $\widehat{\text{ED}}$ is 
the standard sample energy distance estimator defined above. Since $n$ 
is fixed and identical across all method comparisons, $\varepsilon_{s,k} 
= \frac{n}{2}\widehat{\text{ED}}_{s,k}$ is a positive scaling of 
$\widehat{\text{ED}}_{s,k}$, and therefore the rankings based on $\overline{\varepsilon}$ are equivalent to those based on 
$\overline{\widehat{\text{ED}}} = \frac{1}{S}\sum_{s=1}^{S}\frac{1}{K}
\sum_{k=1}^{K}\widehat{\text{ED}}_{s,k}$.

We also compute the squared maximum mean discrepancy, denoted by $\mathrm{MMD}^2$, using the \texttt{kmmd} function from the \texttt{kernlab} package with a Gaussian kernel and automatic bandwidth selection:
\[
\mathrm{MMD}^2
=
\frac{1}{S}\sum_{s=1}^S
\frac{1}{K}\sum_{k=1}^K \mathrm{MMD}_{s,k}^2,
\]
where
\[
\mathrm{MMD}_{s,k}^2
=
\frac{1}{n(n-1)}\sum_{i\neq j} k(X_i^{\mathrm{true}},X_j^{\mathrm{true}})
+
\frac{1}{n(n-1)}\sum_{i\neq j} k(X_{i,s,k}^{\mathrm{imp}},X_{j,s,k}^{\mathrm{imp}})
-
\frac{2}{n^2}\sum_{i,j} k(X_i^{\mathrm{true}},X_{j,s,k}^{\mathrm{imp}}).
\]
Smaller values of all five metrics indicate better imputation performance.

\section{Additional Simulation Results}
\label{app:sim_results}

This section reports additional simulation results that complement the simulation studies in the main paper, including results under higher missingness, MAR settings, varying complete-case proportions, and pointwise evaluation metrics.

\subsection{Results Under 40\% Missingness}

\begin{figure}[t]
    \centering
    \includegraphics[width=\linewidth]{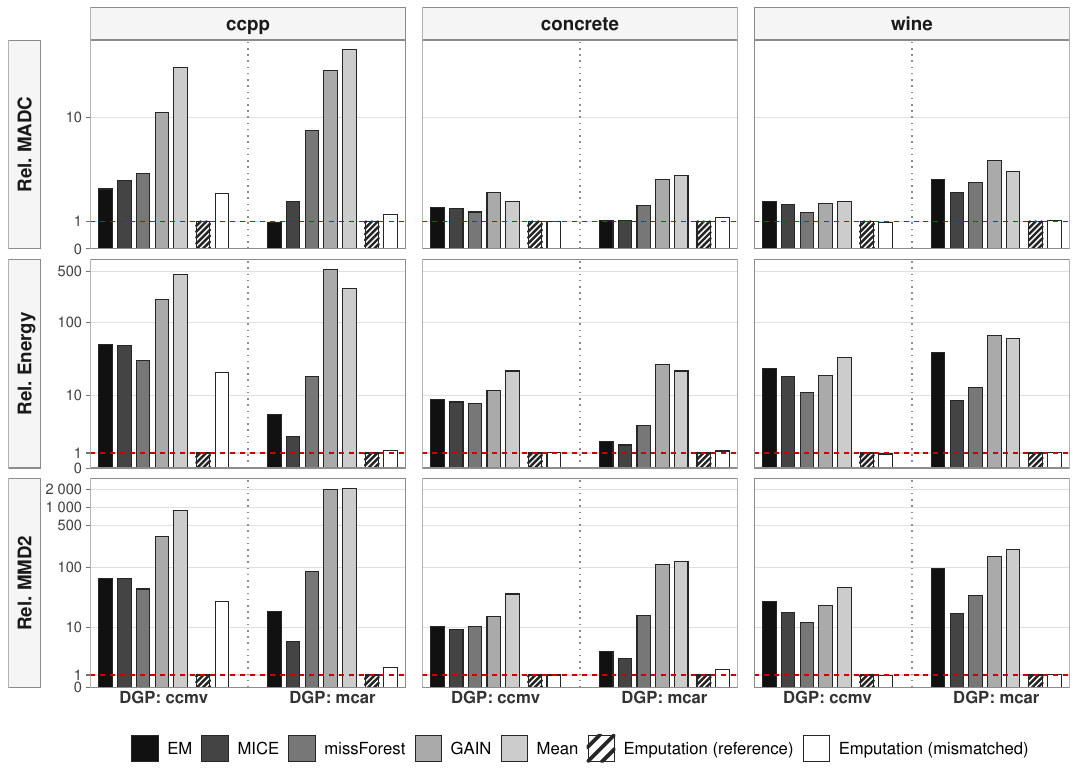}
    \caption{Simulation under 40\% missingness: performance of imputation methods in MADC, energy distance relative to emputation, and $\mathrm{MMD}^2$ relative to emputation. The y-axis is shown on a logarithmic scale to accommodate the large variation across methods. The x-axis shows the missingness-generation mechanism.}
    \label{fig:sim40_distributional}
\end{figure}

\autoref{fig:sim40_distributional} reports the simulation results under 40\% missingness in terms of MADC, energy distance, and $\mathrm{MMD}^2$. The results are consistent with the findings in the main text. 
Emputation achieves the best distributional performance across datasets and missingness mechanisms. 
The matched emputation model generally outperforms the mismatched version, indicating the benefit of aligning the emputation risk with the missingness-generation mechanism, with larger gains observed for CCPP. For Wine, the mismatched version is occasionally slightly better under MCAR emputation, which may be due to a larger training sample.

\subsection{Additional MAR Results}
\begin{figure}[t]
     \centering
     \begin{subfigure}[b]{0.49\linewidth}
         \centering
         \includegraphics[width=\linewidth]{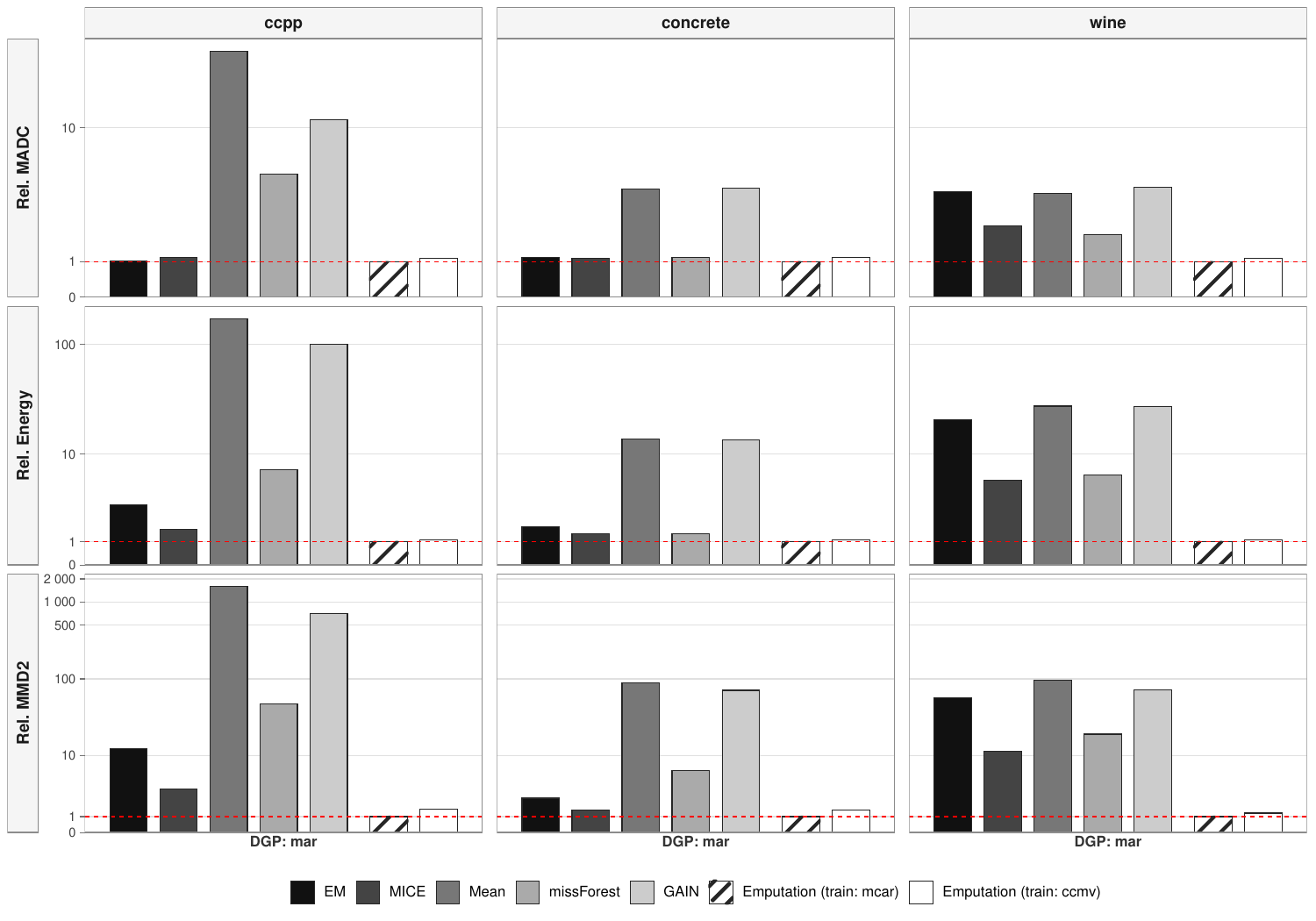}
         \caption{MAR (20\%)}
         \label{fig:mar20_distributional}
     \end{subfigure}
     \hfill
     \begin{subfigure}[b]{0.49\linewidth}
         \centering
         \includegraphics[width=\linewidth]{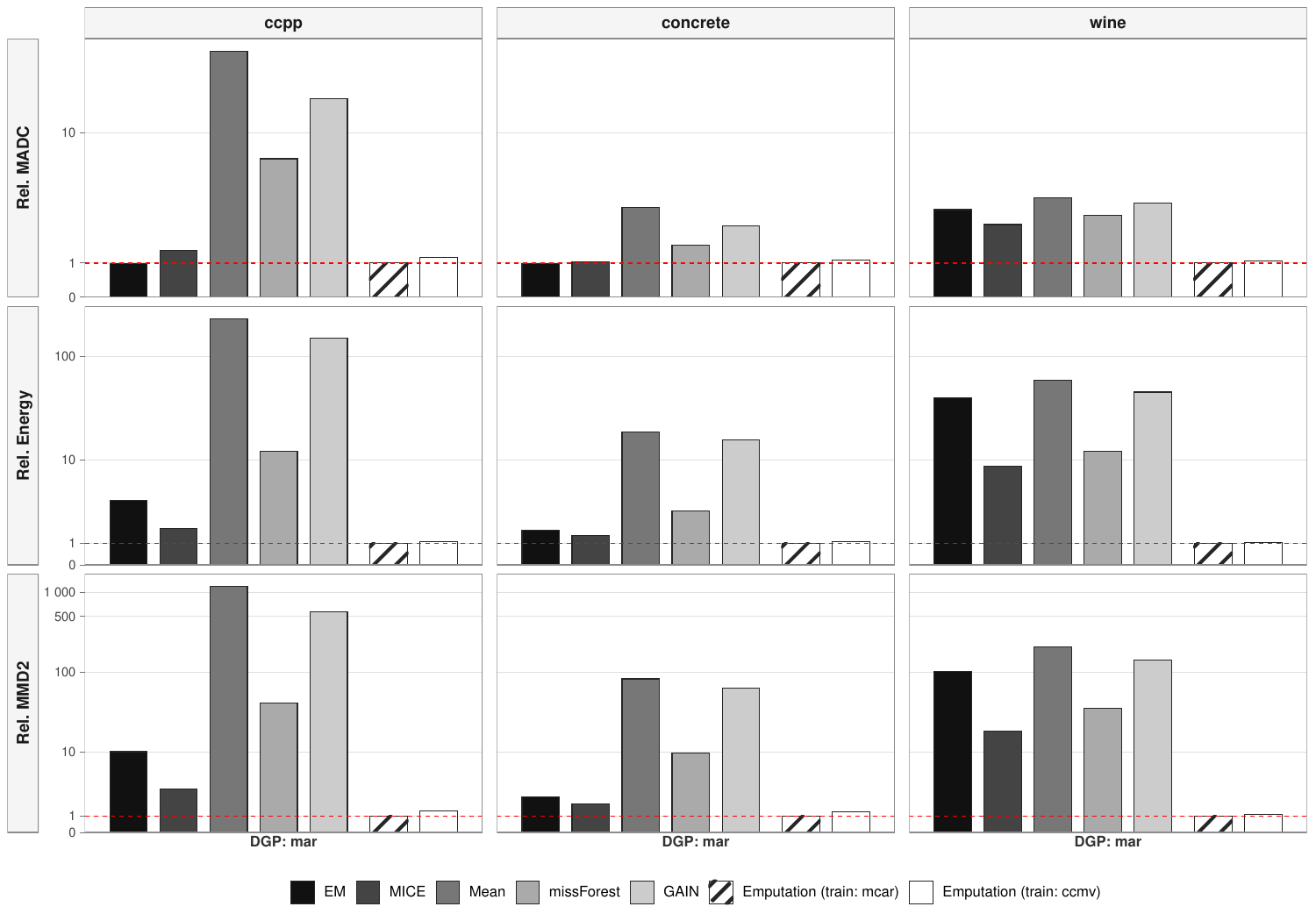}
         \caption{MAR (40\%)}
         \label{fig:mar40_distributional}
     \end{subfigure}
     
     \caption{Simulation under MAR: performance of imputation methods in MADC, energy distance, and $\mathrm{MMD}^2$ (relative to emputation-MCAR). The y-axis is shown on a log scale.}
     \label{fig:mar_comparison}
\end{figure}

\autoref{fig:mar_comparison} shows results when missing data are generated under MAR, with approximately 20\% and 40\% missingness.
Although neither emputation specification matches the MAR mechanism, emputation achieves the best performance in terms of energy distance and MMD$^2$ across datasets. For MADC, emputation is also among the top-performing methods, with MICE and EM showing competitive results on the CCPP and Concrete data. Emputation models trained under MCAR and CCMV assumptions yield similar performance overall. In some cases, the MCAR-trained model performs slightly better, which is likely due to the larger training sample size. Overall, these empirical results suggest that emputation can remain competitive under MAR. Theoretical guarantees for emputation under MAR are discussed as a future direction in the main paper.

\subsection{Complete-Case Proportion}
Since the CCMV emputation risk uses complete cases during training, we examine how the complete-case proportion affects imputation performance in terms of energy distance. We conduct this analysis under MAR, training emputation under both MCAR and CCMV assumptions. We vary the complete-case proportion from 5\% to 60\% and generate missingness under MAR. MICE and EM are included as benchmark methods. Each setting is repeated over 100 independent iterations. The resulting confidence intervals are narrow and thus omitted from the figure for clarity. 

\begin{figure}[t]
    \centering
    \includegraphics[width=\linewidth]{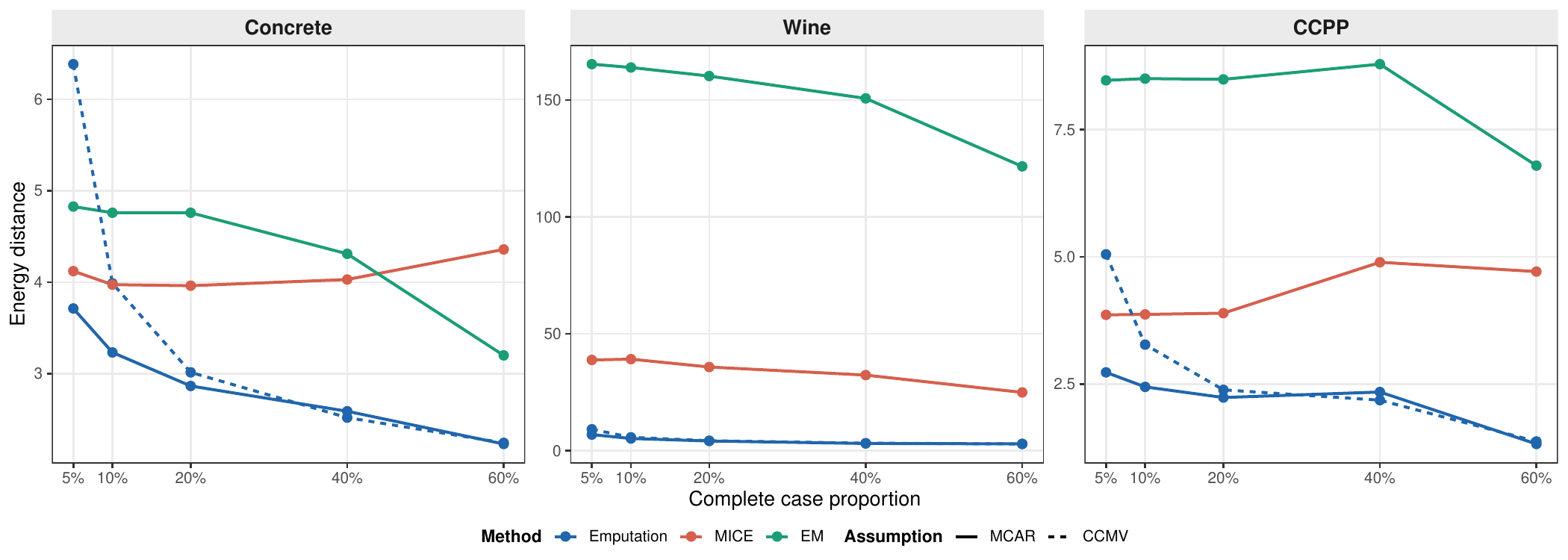}
    \caption{Energy distance as a function of the complete-case proportion. Missingness is generated under MAR.}
    \label{fig:ccprop_combined}
\end{figure}

As shown in \autoref{fig:ccprop_combined}, emputation remains competitive across a range of complete-case proportions and generally outperforms MICE and EM. 
The dependence on the complete-case proportion is most visible for Concrete, where performance improves as more complete cases are available. 
Interestingly, MICE does not necessarily improve monotonically as the complete-case proportion increases, and its energy distance even increases for Concrete and CCPP. This may be due to imputation model misspecification. In our implementation, MICE uses predictive mean matching. When the conditional relationships among variables are strongly nonlinear, as may be the case for Concrete and CCPP, increasing the amount of observed information does not necessarily improve performance in terms of energy distance if the working imputation model remains misspecified. In contrast, emputation uses a flexible neural-network-based model, which may partly explain its superior performance even though the MCAR and CCMV training assumptions are not the true MAR mechanism in this experiment. For Wine, MICE shows a slight decreasing trend as the complete-case proportion increases, which may reflect dataset-specific features, since it has the largest number of variables among the three datasets.
Overall, these simulations suggest that a complete-case proportion of at least 20\% may serve as a useful practical guideline for applying CCMV emputation. 

\subsection{Pointwise Accuracy}
\autoref{tab:pointwise_accuracy_20} and \autoref{tab:pointwise_accuracy_40} report the results of pointwise metrics for 20\% and 40\% missingness, respectively. Overall, emputation and missForest consistently have the smallest RMSE and MAE across all settings. It is not surprising that missForest yields low RMSE and MAE. As a nonparametric imputation approach, missForest is trained by minimizing pointwise accuracy. However, missForest exhibits larger energy distance and MMD$^2$ than emputation in most settings. These results emphasize that strong pointwise accuracy does not necessarily imply distribution alignment, whereas emputation directly targets distributional learning.

\begin{table}[ht]
\centering
\caption{Pointwise accuracy under 20\% missingness simulation. Within each row, the smallest two values for RMSE and MAE are shown in bold. DGP denotes the missing data generation process. Emp-MCAR and Emp-CCMV denote emputation trained with the MCAR and CCMV emputation risks, respectively. Competing methods include Expectation-Maximization (EM), GAIN, missForest, MICE, and mean imputation (Mean).}
\label{tab:pointwise_accuracy_20}
\resizebox{\textwidth}{!}{
\begin{tabular}{llccccccc|ccccccc}
\toprule
& & \multicolumn{7}{c|}{RMSE} & \multicolumn{7}{c}{MAE} \\
\cmidrule(lr){3-9} \cmidrule(lr){10-16}
Data & DGP
& Emp-MCAR & Emp-CCMV & EM & GAIN & MissForest & MICE & Mean
& Emp-MCAR & Emp-CCMV & EM & GAIN & MissForest & MICE & Mean \\
\midrule

\multirow{3}{*}{Concrete}
& MCAR
& \textbf{0.6188} & 0.6293 & 0.7326 & 0.9768 & \textbf{0.5571} & 0.7118 & 1.0013
& \textbf{0.4199} & 0.4289 & 0.5465 & 0.7623 & \textbf{0.3248} & 0.5063 & 0.8115 \\
& CCMV
& \textbf{0.6239} & \textbf{0.6238} & 0.8046 & 1.0759 & 0.7168 & 0.7495 & 1.0385
& \textbf{0.4440} & \textbf{0.4439} & 0.6195 & 0.8559 & 0.5042 & 0.5566 & 0.8784 \\
& MAR
& \textbf{0.6007} & 0.6158 & 0.7161 & 0.9838 & \textbf{0.5320} & 0.6952 & 1.0101
& \textbf{0.4042} & 0.4165 & 0.5296 & 0.7650 & \textbf{0.3045} & 0.4900 & 0.8156 \\
\midrule

\multirow{3}{*}{Wine}
& MCAR
& \textbf{0.7735} & 0.7896 & 0.9577 & 1.0483 & \textbf{0.7123} & 0.9325 & 1.0013
& \textbf{0.5364} & 0.5477 & 0.7165 & 0.7888 & \textbf{0.4756} & 0.6957 & 0.7597 \\
& CCMV
& \textbf{0.7343} & \textbf{0.7339} & 0.9753 & 1.0702 & 0.8219 & 0.9714 & 0.9085
& \textbf{0.5243} & \textbf{0.5233} & 0.7807 & 0.8374 & 0.6078 & 0.7750 & 0.7351 \\
& MAR
& \textbf{0.7614} & 0.7773 & 0.9502 & 1.0312 & \textbf{0.6925} & 0.9243 & 0.9958
& \textbf{0.5294} & 0.5399 & 0.7127 & 0.7790 & \textbf{0.4640} & 0.6920 & 0.7581 \\
\midrule

\multirow{3}{*}{CCPP}
& MCAR
& 0.6071 & \textbf{0.6069} & 0.6421 & 0.8304 & \textbf{0.5433} & 0.6274 & 1.0004
& 0.4281 & \textbf{0.4279} & 0.4680 & 0.6445 & \textbf{0.3643} & 0.4473 & 0.8538 \\
& CCMV
& 0.5659 & \textbf{0.5539} & 0.6247 & 0.8219 & \textbf{0.5530} & 0.6055 & 1.1072
& 0.3931 & \textbf{0.3874} & 0.4518 & 0.6260 & \textbf{0.3601} & 0.4224 & 0.9620 \\
& MAR
& 0.6005 & \textbf{0.6000} & 0.6363 & 0.8270 & \textbf{0.5256} & 0.6201 & 1.0047
& 0.4189 & \textbf{0.4184} & 0.4605 & 0.6406 & \textbf{0.3501} & 0.4378 & 0.8577 \\
\bottomrule
\end{tabular}
}
\end{table}

\begin{table}[ht]
\centering
\caption{Pointwise accuracy under 40\% missingness simulation. Within each row, the smallest two values for RMSE and MAE are shown in bold. DGP denotes the missing data generation process. Emp-MCAR and Emp-CCMV denote emputation trained with the MCAR and CCMV emputation risks, respectively. Competing methods include Expectation-Maximization (EM), GAIN, missForest, MICE, and mean imputation (Mean).}
\label{tab:pointwise_accuracy_40}
\resizebox{\textwidth}{!}{
\begin{tabular}{llccccccc|ccccccc}
\toprule
& & \multicolumn{7}{c|}{RMSE} & \multicolumn{7}{c}{MAE} \\
\cmidrule(lr){3-9} \cmidrule(lr){10-16}
Data & DGP
& Emp-MCAR & Emp-CCMV & EM & GAIN & MissForest & MICE & Mean
& Emp-MCAR & Emp-CCMV & EM & GAIN & MissForest & MICE & Mean \\
\midrule

\multirow{3}{*}{Concrete}
& MCAR
& \textbf{0.7748} & 0.7862 & 0.8696 & 1.0274 & \textbf{0.7849} & 0.8617 & 1.0001
& \textbf{0.5552} & 0.5638 & 0.6703 & 0.8032 & \textbf{0.5189} & 0.6528 & 0.8103 \\
& CCMV
& \textbf{0.7414} & \textbf{0.7421} & 0.9725 & 1.1766 & 0.8523 & 0.9364 & 1.0809
& \textbf{0.5302} & \textbf{0.5307} & 0.7786 & 0.9094 & 0.6196 & 0.7431 & 0.9269 \\
& MAR
& \textbf{0.7904} & 0.7954 & 0.8699 & 0.9664 & \textbf{0.7671} & 0.8631 & 1.0077
& \textbf{0.5784} & 0.5811 & 0.6704 & 0.7551 & \textbf{0.5032} & 0.6539 & 0.8143 \\
\midrule

\multirow{3}{*}{Wine}
& MCAR
& \textbf{0.8384} & 0.8697 & 1.0080 & 1.0819 & \textbf{0.8674} & 0.9672 & 1.0012
& \textbf{0.5901} & 0.6243 & 0.7558 & 0.8037 & \textbf{0.6107} & 0.7260 & 0.7598 \\
& CCMV
& \textbf{0.7749} & \textbf{0.7830} & 1.0399 & 1.0193 & 0.8680 & 1.0065 & 1.0127
& \textbf{0.5610} & \textbf{0.5645} & 0.8390 & 0.8083 & 0.6508 & 0.8160 & 0.8429 \\
& MAR
& \textbf{0.8537} & 0.8572 & 1.0003 & 0.9905 & \textbf{0.8475} & 0.9601 & 0.9972
& \textbf{0.6137} & 0.6153 & 0.7507 & 0.7318 & \textbf{0.5970} & 0.7218 & 0.7586 \\
\midrule

\multirow{3}{*}{CCPP}
& MCAR
& \textbf{0.6937} & \textbf{0.6943} & 0.7215 & 1.1320 & 0.7101 & 0.7164 & 1.0001
& \textbf{0.5041} & 0.5046 & 0.5384 & 0.8806 & \textbf{0.4991} & 0.5280 & 0.8535 \\
& CCMV
& \textbf{0.6467} & \textbf{0.5854} & 0.7349 & 0.9880 & 0.6693 & 0.7272 & 1.2418
& \textbf{0.4467} & \textbf{0.4159} & 0.5303 & 0.7626 & 0.4492 & 0.5134 & 1.0749 \\
& MAR
& \textbf{0.6835} & \textbf{0.6834} & 0.7048 & 0.8750 & 0.6917 & 0.6975 & 1.0036
& \textbf{0.4942} & 0.4945 & 0.5224 & 0.6751 & \textbf{0.4820} & 0.5090 & 0.8568 \\
\bottomrule
\end{tabular}
}
\end{table}

\section{NACC Data and Sensitivity Analysis}
\label{app:nacc_data}
\autoref{tab::nacc_vardescription} summarizes participant characteristics by cognitive status. On average, impaired participants are younger at death, while education and race distributions are similar across groups. Genetically, APOE $\varepsilon$4 carrier status is more common among impaired participants, whereas APOE $\varepsilon$2 carrier status is more common among cognitively normal participants. ADNC also varies by cognitive status, with high ADNC much more frequent in the impaired group and low or absent ADNC more common among cognitively normal participants.

\begin{table}[ht]
\centering
\caption{Demographic and genetic characteristics by cognitive status. Statistics are presented as mean (standard deviation) for age and education and count (\%) for the other variables.}
\label{tab::nacc_vardescription}
\begin{threeparttable}
\renewcommand{\arraystretch}{0.85}
\setlength{\tabcolsep}{3pt}       
\footnotesize         
\begin{tabular}{lccc}
\toprule
& Overall & Normal (CDR = 0) & Impaired (CDR $>$ 0) \\
& (N = 4906) & (N = 645) & (N = 4261) \\
\midrule

\textbf{Age at death, years} & 81 (11) & 86 (9) & 80 (11) \\

\textbf{Sex} & & & \\
\quad Male & 2680 (55\%) & 285 (44\%) & 2395 (56\%) \\
\quad Female & 2226 (45\%) & 360 (56\%) & 1866 (44\%) \\

\textbf{Race} & & & \\
\quad White & 4641 (95\%) & 607 (94\%) & 4034 (95\%) \\
\quad Non-White & 265 (5.4\%) & 38 (5.8\%) & 227 (5.3\%) \\

\textbf{Education, years} & 15 (3) & 16 (3) & 15 (3) \\

\textbf{APOE $\varepsilon$4 carrier} & 2075 (42\%) & 129 (20\%) & 1946 (46\%) \\

\textbf{APOE $\varepsilon$2 carrier} & 571 (12\%) & 110 (17\%) & 461 (11\%) \\

\textbf{ADNC} & & & \\
\quad Not AD & 522 (10\%) & 124 (19\%) & 398 (9.2\%) \\
\quad Low & 1429 (29\%) & 375 (58\%) & 1054 (25\%) \\
\quad Intermediate & 912 (19\%) & 130 (20\%) & 782 (18\%) \\
\quad High & 2043 (42\%) & 16 (2.5\%) & 2027 (48\%) \\

\bottomrule
\end{tabular}
\end{threeparttable}
\end{table}

\subsection{Tree Graph} 
\begin{figure}[ht]
\centering
\begin{tikzpicture}[
	every node/.style={font=\Large},
	edge from parent/.style={
		draw,
		-{Stealth[length=2mm]},
		thin
	},
	level 1/.style={sibling distance=2.8cm, level distance=1.5cm},
	level 2/.style={sibling distance=2.0cm, level distance=1.5cm},
	level 3/.style={sibling distance=2.0cm, level distance=1.5cm}
	]
	
	\node {111}
	child { node {011} }
	child { node {101} }
	child { node {110}
		child { node {010}
			child { node {000} }
		}
		child { node {100} }
	}
	child { node {001} };
	
\end{tikzpicture}
\caption{Tree graph for the NACC data analysis obtained by parent-based alignment. Each node represents a response pattern, where the three 
	binary digits correspond to \texttt{NACCBMI}, \texttt{BPSYS}, and \texttt{NACCGDS}, respectively.}
\label{fig::tree}
\end{figure}
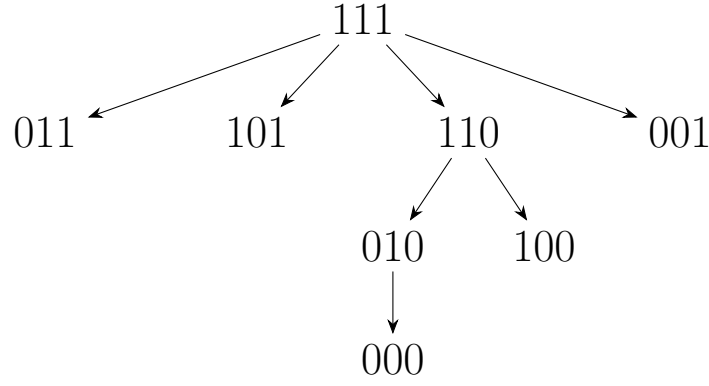

We select the tree graph using the parent-based alignment approach in \cite{suen2026modelingmultivariatemissingnesstree}. For each non-complete pattern $r\neq 1_d$, the parent $s$ is chosen as the one whose observed data distribution $p(x_r|R=s)$ most closely matches $p(x_r|R=r)$, as measured by the energy distance computed on the variables commonly observed among \texttt{NACCBMI}, \texttt{BPSYS}, and \texttt{NACCGDS} together with the always-observed variables. Continuous variables (\texttt{AGE}, \texttt{EDUC}, \texttt{NACCBMI}, \texttt{BPSYS}, \texttt{NACCGDS}) are standardized prior to distance computation, while categorical variables (\texttt{SEX}, \texttt{RACE}, \texttt{NACCAPOE}, \texttt{NACCDIUR}, \texttt{NPADNC}, \texttt{CDRGLOB}) are kept on their original scales, so that no single variable dominates the energy distance. The resulting tree graph is shown in \autoref{fig::tree}.

We further apply the tree-graph emputation to the NACC data using 500 bootstrap replicates. The training parameters used for tree-graph emputation are the same as those reported in the main analysis under the MCAR and CCMV assumptions. The imputation procedure,  as detailed in the main paper, proceeds sequentially along the tree. For each incomplete pattern, missing variables are imputed step by step from the observed pattern toward the root node. For instance, consider pattern $000$, where \texttt{NACCBMI}, \texttt{BPSYS}, and \texttt{NACCGDS} are all missing. According to the selected tree graph, the path connecting $000$ to the root is $000 \leftarrow 010 \leftarrow 110 \leftarrow 111$. Thus, the imputation is performed sequentially along this path. \texttt{BPSYS} is first imputed to move from $000$ to $010$, then \texttt{NACCBMI} is imputed to move from $010$ to $110$, and finally \texttt{NACCGDS} is imputed to reach the complete-case pattern $111$.

\begin{figure}
    \centering
    \includegraphics[width=\linewidth]{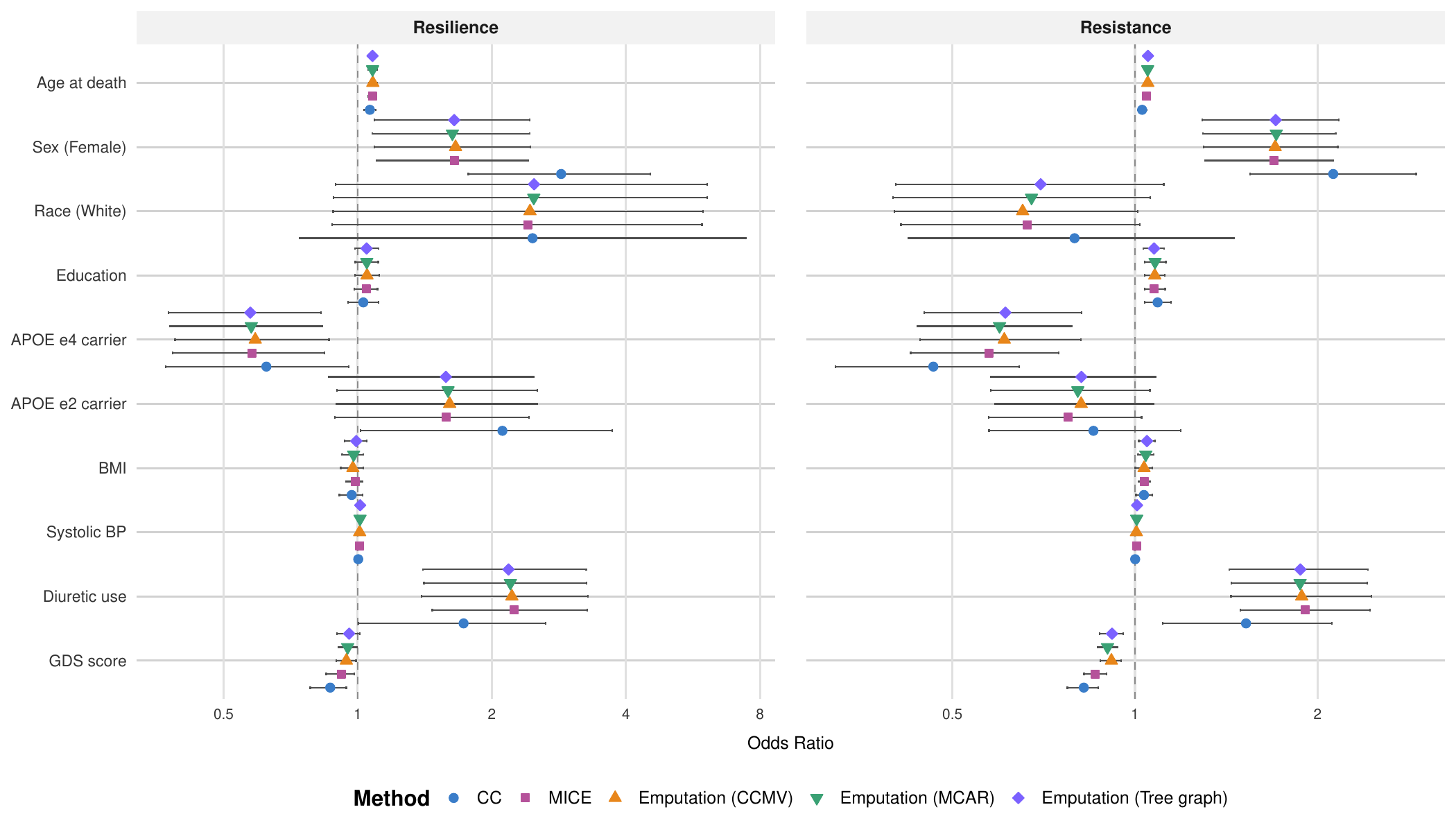}
    \caption{Odds ratios (ORs) and 95\% confidence intervals for predictors of cognitive resilience (left) and cognitive resistance (right) under five missing-data analyses: complete-case analysis (CC), multiple imputation by chained equations (MICE), emputation under MCAR, emputation under CCMV, and emputation under tree graph.}
    \label{fig::nacc_tree}
\end{figure}

As shown in \autoref{fig::nacc_tree}, the estimates from tree-graph emputation are  consistent with those obtained under CCMV and MCAR assumptions. For both the resilience and resistance models, the odds ratio estimates and 95\% confidence intervals are similar across the three emputation assumptions and are also aligned with MICE. It is interesting that complete-case analysis yields very different estimates from emputation for several covariates such as sex, diuretic use, and GDS score. Although MCAR is a strong working assumption, emputation uses incomplete observations during training and thereby exploits much more observed information than complete-case analysis. In addition, emputation uses a flexible neural-network-based distributional model, which may better capture nonlinear dependence among clinical and demographic variables. Overall, \autoref{fig::nacc_tree} suggests that the tree-graph analysis supports the scientific conclusions reported in the main paper and also suggests that complete-case analysis can yield different results when incomplete observations are discarded.

\subsection{Sensitivity Analysis}
Emputation accommodates a variety of missing data assumptions. It is therefore important to assess the sensitivity of downstream conclusions to this choice. We have compared estimated odds ratios across MCAR, CCMV, and tree graph emputation risks, which examine sensitivity to different identifying assumptions on the extrapolation density. Here, we further assess departures from the CCMV assumption using an exponential tilting scheme. For a sensitivity parameter $\rho$, the tilted extrapolation density is 
\[
p_{\rho}(x_{\bar r}|x_r,R=r)\propto p(x_{\bar r}|x_r,R=1_d) \cdot \exp(-\rho \|x_{\bar r}-\mu_{\bar r}\|^2)
\]
where $\rho = 0$ recovers CCMV, and $\mu_{\bar r} = \mathbb{E}[X_{\bar r}|X_r, R=1_d]$ is the conditional mean under the CCMV extrapolation density. In practice, the tilting is operationalized in the standardized logit-transformed space to ensure scale invariance across variables with different supports, specifically BMI $\in [10, 100]$, systolic blood pressure $\in [70, 230]$, and GDS score $\in [0, 15]$. The quantity $\mu_{\bar r}$ is estimated as the Monte Carlo mean of $M$ candidate draws from the trained emputation model. We obtain samples from the tilted density via sampling-importance resampling, where each candidate draw receives importance weight $w^{(m)} \propto \exp\left(-\rho\|x_{\bar r}^{(m)} - \mu_{\bar r}\|^2\right)$,
and $K$ draws are resampled for each subject according to the normalized weights $\tilde{w}^{(m)} = w^{(m)}/\sum_m w^{(m)}$. The resampled draws are then transformed back to the original scale via destandardization and inverse logit. These draws serve as multiple imputations, and odds ratios are pooled across the $K$ completed datasets. Bootstrap confidence intervals are constructed across $B = 500$ replicates.

\begin{figure}
    \centering
    \includegraphics[width=\linewidth]{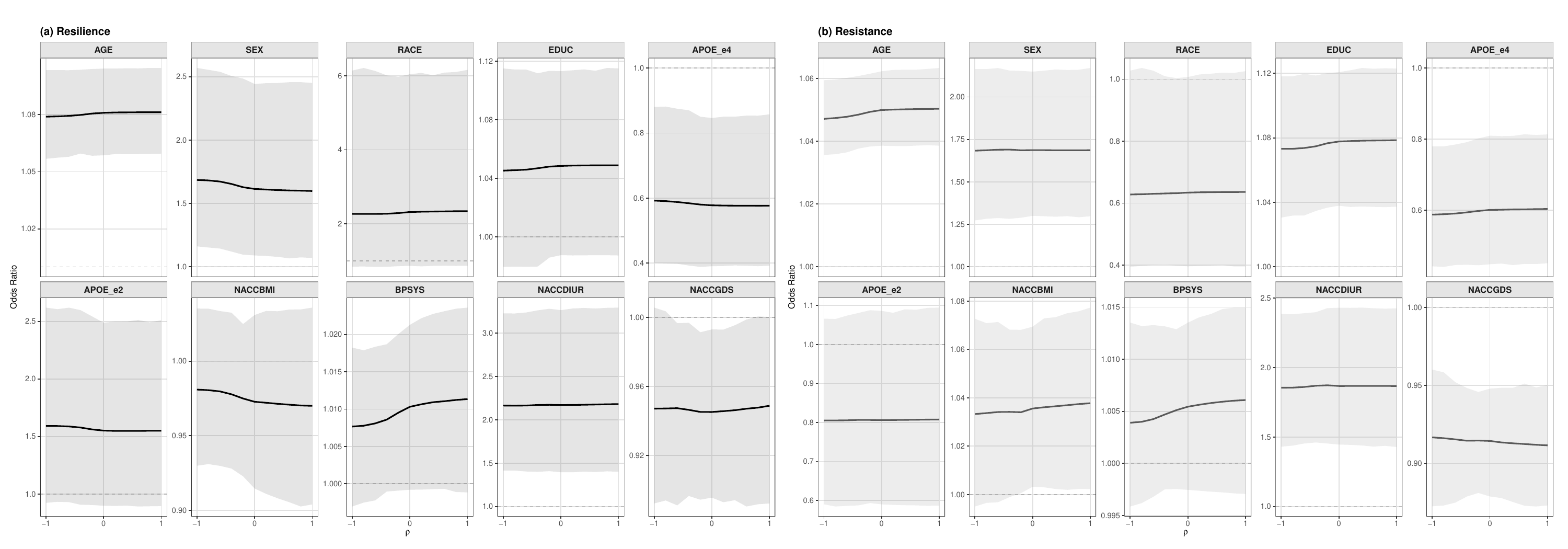}
    \caption{Sensitivity analysis: estimated odds ratios across values of the sensitivity parameter $\rho$ for (a) resilience model and (b) resistance model.}
    \label{fig:sensitivity}
\end{figure}

\autoref{fig:sensitivity} shows that the associations for age, sex, APOE $\varepsilon$4 carrier status, and diuretic use are stable in both models, with both the direction and statistical significance robust to departures from CCMV. BMI in the resistance model and GDS in the resilience model are more sensitive, with statistical significance not consistently preserved across different values of $\rho$. These associations should therefore be interpreted more cautiously.
Overall, the sensitivity analysis supports the robustness of the main findings and suggests that complete-case analysis may overstate significance because it relies on a selected complete-case sample.

\section{Proofs}
\label{app:proofs}

\subsection{Proof of Theorem 3.1}
The MCAR emputation risk can be written as
\begin{equation*}
\begin{aligned}
\E[\hat{\mathcal{L}}(f)]
&= \E\Bigg(\frac{1}{n}\sum_{i=1}^n\Bigg[
\sum_{r<\bR_i} \frac{1}{|\bR_i-r|}
\Bigg\{
\frac{1}{B}\sum_{b=1}^B \left\|
\bX_{i,\bR_i-r} - [f(\bX_{i,r}, \epsilon_{i,\bar{r}, b}, r)]_{\bR_i-r}\right\| \\
&\quad -\frac{1}{2B(B-1)}\sum_{b, b'=1}^B \left\|[f(\bX_{i,r}, \epsilon_{i,\bar{r}, b}, r)]_{\bR_i-r}  - [f(\bX_{i,r}, \epsilon_{i,\bar{r}, b'},r)]_{\bR_i-r}\right\|\Bigg\} \Bigg]\Bigg) \\
&=\E_{X_R,R,\epsilon}\Bigg(\sum_{r<R} \frac{1}{|R-r|}\Bigg\{
\frac{1}{B}\sum_{b=1}^B \left\|
X_{R-r} - [f(X_{r}, \epsilon_{\bar{r}, b}, r)]_{R-r}\right\| \\
&\quad -\frac{1}{2B(B-1)}\sum_{b, b'=1}^B \left\|[f(X_{r}, \epsilon_{\bar{r}, b}, r)]_{R-r} - [f(X_{r}, \epsilon_{\bar{r}, b'}, r)]_{R-r}\right\| \Bigg\}
\Bigg)\\
&= \E_{X_R,R}\Bigg(\sum_{r<R} \frac{1}{|R-r|}\E_{\epsilon}\left[\norm{X_{R-r} - [f(X_{r}, \epsilon_{\bar{r}, b}, r)]_{R-r}}\right] \\
&\quad - \frac{1}{2}\E_{\epsilon}\norm{[f(X_{r}, \epsilon_{\bar{r}, b}, r)]_{R-r} - [f(X_{r}, \epsilon_{\bar{r}, b'}, r)]_{R-r}}
\Bigg)\\
&=\E_{X_R,R}\Bigg(\sum_{r<R} \frac{1}{|R-r|} [-\text{ES}(p_f(\cdot|x_r,r), X_{R-r})]\Bigg) \\
&=\E_{R}\Bigg(\sum_{r<R} \frac{1}{|R-r|} \E_{X_r|R}\E_{X_{R-r}\sim p(\cdot|x_r,R)}[-\text{ES}(p_f(\cdot|x_r,r), X_{R-r})]\Bigg) \\
&= \E_{R}\Bigg(\sum_{r<R} \frac{1}{|R-r|} \E_{X_r}\E_{X_{R-r}\sim p(\cdot|x_r)}[-\text{ES}(p_f(\cdot|x_r,r), X_{R-r})]\Bigg), 
\end{aligned}
\end{equation*}
where the last equality uses the property  of identified extrapolation densities under MCAR, \emph{i.e.}, $p(\cdot|x_r,R) = p(\cdot| x_r)$ for any $R\in \mathcal{R}\cup \{1_d\}$.
Because the energy score (ES) is a strictly proper scoring rule, the innermost expectation 
\[
\E_{X_{R-r}\sim p(\cdot|x_r)}[-\text{ES}(p_f(\cdot|x_r,r), X_{R-r})] 
\]
is minimized if and only if, for any $s\in\mathcal{R}\cup \{1_d\}$ satisfying $s>r$,
\[
p_{f^*}(x_{s-r}|x_r,r) = p(x_{s-r}|x_r).
\]
Therefore, the density of $\widehat{X}_{\bar r}$ satisfies $p_{f^*}(x_{\bar r}|x_r,r) = p(x_{\bar r}|x_r)$ and recovers the true extrapolation density under MCAR.

\subsection{Proof of Theorem 3.2}
The CCMV population risk can be expressed as  
\begin{equation*}
\begin{aligned}
\E[\hat{\mathcal{L}}(f)] 
&= \E\Bigg(\frac{1}{n}\sum_{i=1}^n\Bigg[
\sum_{r\in\mathcal{R}} I(\bR_i=1_d) \frac{1}{|\bar r|}
\Bigg\{
\frac{1}{B}\sum_{b=1}^B \left\|
\bX_{i,\bar r} - [f(\bX_{i,r}, \epsilon_{i,\bar{r}, b}, r)]_{\bar r}\right\| \\
&\quad-\frac{1}{2B(B-1)}\sum_{b, b'=1}^B \left\|[f(\bX_{i,r}, \epsilon_{i,\bar{r}, b}, r)]_{\bar r}  - [f(\bX_{i,r}, \epsilon_{i,\bar{r}, b'},r)]_{\bar r}\right\| 
\Bigg\} \Bigg]\Bigg) \\
&=\E_{X_R,R,\epsilon}\Bigg(
\sum_{r\in\mathcal{R}} I(R=1_d) \frac{1}{|\bar r|}
\Bigg\{
\frac{1}{B}\sum_{b=1}^B \left\|
X_{\bar r} - [f(X_{r}, \epsilon_{\bar{r}, b}, r)]_{\bar r}\right\| \\
&\quad-\frac{1}{2B(B-1)}\sum_{b, b'=1}^B \left\|[f(X_r, \epsilon_{\bar{r}, b}, r)]_{\bar r}  - [f(X_r, \epsilon_{\bar{r}, b'},r)]_{\bar r}\right\| 
\Bigg\} \Bigg)\\
&=\E_{X_R,R}\Bigg(
\sum_{r\in\mathcal{R}} I(R=1_d) \frac{1}{|\bar r|}
\Bigg\{\E_{\epsilon}\left[\norm{X_{\bar r} - [f(X_{r}, \epsilon_{\bar{r}, b}, r)]_{\bar r}}\right] \\
&\quad - \frac{1}{2}\E_{\epsilon}\norm{[f(X_{r}, \epsilon_{\bar{r}, b}, r)]_{\bar r} - [f(X_{r}, \epsilon_{\bar{r}, b'}, r)]_{\bar r}} \Bigg\}\Bigg)\\
&= \E_{X_R,R}\Bigg(
\sum_{r\in\mathcal{R}} I(R=1_d) \frac{1}{|\bar r|} [-\text{ES}(p_{f}(\cdot|x_r,r), X_{\bar r})]\Bigg)\\
&= p(R=1_d)\sum_{r\in\mathcal{R}}\frac{1}{|\bar r|}\E_{X_r|R=1_d}\E_{X_{\bar r}\sim p(\cdot|x_r, R=1_d)}[-\text{ES}(p_{f}(\cdot|x_r,r), X_{\bar r})].
\end{aligned}
\end{equation*}
Because the energy score is a strictly proper scoring rule, the innermost expectation 
\[
\E_{X_{\bar r}\sim p(\cdot|x_r, R=1_d)}[-\text{ES}(p_{f}(\cdot|x_r,r), X_{\bar r})]
\]
is minimized if and only if \[
p_{f^*}(x_{\bar r}|x_r,r) = p(x_{\bar r}|x_r, R=1_d).
\]

\subsection{Proof of Theorem 3.3}
Similar to the derivation above, the tree-graph emputation risk can be written as
\begin{equation*}
\begin{aligned}
\mathbb{E}[\hat{L}_G(f)]
&= \mathbb{E}\Bigg[\frac{1}{n} \sum_{i=1}^n \sum_{r\in\mathcal{R}}
\frac{I(\bR_i = {\sf PA}(r))}{|{\sf PA}(r)-r|}
\Bigg\{
    \frac{1}{B} \sum_{b=1}^B \left\| \bX_{i, {\sf PA}(r)-r} - 
    [f(\bX_{i,r}, \epsilon_{i,\bar{r},b}, r)]_{{\sf PA}(r)-r} \right\|\\
&\quad - \frac{1}{2B(B-1)} \sum_{b \neq b'} \left\| 
    [f(\bX_{i,r}, \epsilon_{i,\bar{r},b}, r)]_{{\sf PA}(r)-r}
    - [f(\bX_{i,r}, \epsilon_{i,\bar{r},b'}, r)]_{{\sf PA}(r)-r} \right\|
\Bigg\}\Bigg]\\
&=\mathbb{E}\Bigg[\sum_{r\in\mathcal{R}} \frac{I(R = {\sf PA}(r))}{|{\sf PA}(r)-r|}
\Bigg\{\E_{\epsilon}\left[\norm{X_{{\sf PA}(r)-r} - [f(X_{r}, \epsilon_{\bar{r}, b}, r)]_{{\sf PA}(r)-r}}\right] \\
&\quad - \frac{1}{2}\E_{\epsilon}\norm{[f(X_{r}, \epsilon_{\bar{r}, b}, r)]_{{\sf PA}(r)-r} - [f(X_{r}, \epsilon_{\bar{r}, b'}, r)]_{{\sf PA}(r)-r}} \Bigg\}
\Bigg]\\
&= \sum_{r\in\mathcal{R}}
\frac{p(R = {\sf PA}(r))}{|{\sf PA}(r)-r|}
    \mathbb{E}_{X_{r} | R = {\sf PA}(r)}
    \mathbb{E}_{X_{{\sf PA}(r)-r} \sim p(\cdot | x_{r},R = {\sf PA}(r))}[-\mathrm{ES}(p_f(\cdot | x_{r}, r), X_{{\sf PA}(r)-r})].
\end{aligned}
\end{equation*}
For each $r\in\mathcal{R}$, the expectation $\mathbb{E}_{X_{{\sf PA}(r)-r} \sim p(\cdot | x_{r}, R={\sf PA}(r))}[-\mathrm{ES}\left(p_f(\cdot | x_{r}, r), X_{{\sf PA}(r)-r}\right)]$ is minimized if and only if
\begin{equation*}
    p_{f^*}(x_{{\sf PA}(r)-r} \mid x_{r}, r) = p(x_{{\sf PA}(r)-r} \mid x_{r}, R = {\sf PA}(r)).
\end{equation*}

\subsection{Proof of Theorem 4.1}
The m-ACMV emputation risk can be written as
\begin{equation*}
\begin{aligned}
\E[\hat{\mathcal{L}}(g)] 
&= \E\Bigg[
\frac{1}{n}\sum_{i=1}^n
\sum_{s=1}^{d-1}
I(\bT_i >s)
\Bigg\{ 
\frac{1}{B}\sum_{b=1}^B \left|
\bX_{i,s+1}  - g(\bX_{i,\leq s}, \epsilon_{i, > s, b}, s)\right| \\
& \quad -\frac{1}{2B(B-1)}\sum_{b\neq b'} \left|g(\bX_{i,\leq s}, \epsilon_{i, > s, b}, s)  - g(\bX_{i,\leq s}, \epsilon_{i, > s, b'}, s)\right| \Bigg\}
\Bigg]\\
&=\E \Bigg[\sum_{s=1}^{d-1}I(T>s)\Bigg(
\E_{\epsilon}\left|
X_{s+1}- g(X_{\leq s}, \epsilon_{>s, b}, s)\right| \\
& \quad -\frac{1}{2}\E_{\epsilon} \left|g(X_{\leq s}, \epsilon_{>s, b}, s)-g(X_{\leq s}, \epsilon_{>s, b'}, s)\right| \Bigg)
\Bigg]\\
&= \E \Bigg[\sum_{s=1}^{d-1}I(T>s)\{-\text{ES}(p_{g}(\cdot|x_{\leq s},s), X_{s+1})\}\Bigg]\\
&=\sum_{s=1}^{d-1}p(T>s)\E_{X_{\leq s}|T>s}\E_{X_{s+1}\sim p(\cdot|x_{\leq s}, T>s)}[-\text{ES}(p_{g}(\cdot|x_{\leq s},s), X_{s+1})]\\
\end{aligned}
\end{equation*}
Since the energy score is strictly proper, the inner expectation \[
\E_{X_{s+1}\sim p(\cdot|x_{\leq s}, T>s)} [-\text{ES}(p_{g}(\cdot|x_{\leq s},s), X_{s+1})]
\]
is minimized if and only if
\[
p_{g^*}(x_{s+1}|x_{\leq s},T=s) = p(x_{s+1}|x_{\leq s}, T\geq s+1).
\]
That is, we have shown that the density of $\hat X_{s+1}$ satisfies 
$p_{g^*}(x_{s+1}|x_{\leq s}, T = s) = p(x_{s+1}|x_{\leq s}, T \geq s+1)$, which recovers the desired extrapolation density under m-ACMV.
\end{document}